\documentclass[letter,11pt]{article}
\usepackage{amsmath}
\usepackage{color}
\usepackage{graphicx}
\usepackage{amssymb}
\usepackage{float}
\usepackage{scrtime}
\usepackage{fancyhdr}
\usepackage{subfigure}
\usepackage{authblk}
\usepackage{lipsum}
\usepackage{rotating}
\usepackage{floatpag}
\usepackage{varioref}
\usepackage{slashed}
\usepackage{enumerate}

\usepackage[colorlinks=true
,urlcolor=blue
,anchorcolor=blue
,citecolor=blue
,filecolor=blue
,linkcolor=black
,menucolor=blue
,pagecolor=blue
,linktocpage=true
]{hyperref}

\usepackage[numbers,sort&compress]{natbib}

\setlipsumdefault{131}

\newcommand{\maton}{\texttt{maton}}
\newcommand{\superiso}{\texttt{SuperIso}}
\newcommand{\susyflavor}{\texttt{susy\_flavor}}
\newcommand{\nufit}{\texttt{NuFIT}}

\newcommand{\SO}[1]{\ensuremath{\mathrm{SO}(#1)}}
\newcommand{\SU}[1]{\ensuremath{\mathrm{SU}(#1)}}
\newcommand{\U}[1]{\ensuremath{\mathrm{U}(#1)}}

\newcommand{\bs}[1]{\ensuremath{\boldsymbol{#1}}}
\newcommand{\bsb}[1]{\ensuremath{\boldsymbol{\overline{#1}}}}

\labelformat{section}{Section #1}
\labelformat{subsection}{Section #1}
\labelformat{equation}{Eq.~(#1)}
\labelformat{figure}{Fig.~#1}
\labelformat{subfigure}{Fig.~\thefigure#1}
\labelformat{table}{Tab.~#1}

\def\a{\alpha}
\def\b{\beta}
\def\p{\phi_a}
\def\ptil{\tilde\phi_a}
\def\be{\begin{equation}}
\def\ee{\end{equation}}
\def\bea{\begin{eqnarray}}
\def\eea{\end{eqnarray}}

\graphicspath{{./figures/}}

\begin{document}

\floatpagestyle{plain}

\pagenumbering{roman}

\renewcommand{\headrulewidth}{0pt}
\rhead{
OHSTPY-HEP-T-12-004\\
LPSC-12329
}
\fancyfoot{}

\title{\huge \bf{Yukawa Unification Predictions for the LHC}}

\author[$\dag$]{Archana Anandakrishnan}
\author[$\dag$]{Stuart Raby}
\author[$^\ddag$]{Ak\i{}n Wingerter}

\affil[$\dag$]{\em Department of Physics, The Ohio State University,\newline
191 W.~Woodruff Ave, Columbus, OH 43210, USA \enspace\enspace\enspace\enspace \medskip}

\affil[$^\ddag$]{\em Laboratoire de Physique Subatomique et de Cosmologie,\newline
UJF Grenoble 1, CNRS/IN2P3, INPG,\newline
53 Avenue des Martyrs, F-38026 Grenoble, France\enspace\enspace\enspace\enspace\enspace}

\maketitle
\thispagestyle{fancy}

\begin{abstract}\normalsize\parindent 0pt\parskip 5pt
This paper is divided into two parts. In the first part we analyze
the consequences, for the LHC, of gauge and third family Yukawa coupling
unification with a particular set of boundary conditions defined at
the GUT scale.  We perform a global $\chi^2$ analysis including the
observables $M_W, M_Z, G_F, \alpha_{em}^{-1},$ $\alpha_s(M_Z), M_t,
m_b(m_b), M_\tau, BR(B \rightarrow X_s \gamma), BR(B_s \rightarrow \mu^+ \mu^-)$ and $M_{h}$.  The fit is performed in
the MSSM in terms of 9 GUT scale parameters, while $\tan\beta$ and
$\mu$ are fixed at the weak scale.  Good fits suggest an upper bound
on the gluino mass, $M_{\tilde g} \lesssim 2$ TeV.  This constraint comes predominantly from
fitting the bottom quark and Higgs masses (assuming a 125 GeV Higgs).
Gluinos should be visible at the LHC in the 14 TeV run but they cannot be described
by the typical simplified models. This is because the branching ratios
for $\tilde g \rightarrow t \bar t \ \tilde \chi^0_{1,2}$, $b \bar b \ \tilde \chi^0_{1,2}$, $t \bar b \ \tilde \chi^-_{1,2}$, $b \bar t \ \tilde \chi^+_{1,2}$, $g \ \tilde \chi^0_{1,2,3,4}$ are comparable.  Stops and sbottoms may also be visible.
Charginos and neutralinos can be light with the LSP predominantly bino-like. In the second part of the paper we analyze a complete three family model and discuss the quality of the global $\chi^2$ fits and the
differences between the third family analysis and the full three
family analysis for overlapping observables.  We note that the light Higgs in our model couples to matter
like the Standard Model Higgs.  Any deviation from this would rule out this model.
\end{abstract}

\clearpage
\newpage

\pagenumbering{arabic}

\section{Introduction}

Gauge coupling unification in supersymmetric grand unified theories (SUSY GUTs) \cite{Dimopoulos:1981yj,
Dimopoulos:1981zb, Ibanez:1981yh, Sakai:1981gr, Einhorn:1981sx, Marciano:1981un}  provides an experimental hint for
low energy SUSY.   However, it does not significantly constrain the spectrum of supersymmetric particles.  On the
other hand, it has been observed that Yukawa coupling unification for the third generation of quarks and leptons
in models, such as $\SO{10}$ or $\SU{4}_c \times \SU{2}_L \times \SU{2}_R$, can place significant constraints on the SUSY spectrum in order
to fit the top, bottom and tau masses \cite{Blazek:2001sb,Baer:2001yy,Blazek:2002ta,Tobe:2003bc,Auto:2003ys}.   These constraints depend on the particular boundary conditions for sparticle masses chosen at the GUT scale (see for example, \cite{Blazek:2002ta,Baer:2009ie,Badziak:2011wm},  which consider different GUT scale
boundary conditions).   In light of the present success of the LHC with the possible observation of the Higgs boson with mass of order 125 GeV
and significant lower bounds on gluino and squark masses, it is a perfect time to review the viability of the constraints on the sparticle spectrum resulting from gauge and third generation Yukawa coupling unification.\footnote{For other analyses in this direction, see \cite{Baer:2008xc,Baer:2012cp}.}
This is what we do in this paper.  In part one of the paper, we perform a global $\chi^2$ analysis assuming $\SO{10}$ boundary conditions for sparticle masses and non-universal Higgs masses,  which we have called ``just so Higgs splitting."   We fit the observables,  $M_W, M_Z, G_F, \alpha_{em}^{-1},$ $\alpha_s(M_Z),
M_t, m_b(m_b), M_\tau, BR(B \rightarrow X_s \gamma),  BR(B_s \rightarrow \mu^+ \mu^-)$ and $M_{h}$ in terms of 11 arbitrary parameters.   These fits then place significant constraints on the gluino mass.

In the second part of the paper we study a complete three family model of quark and lepton Yukawa couplings at the GUT scale \cite{Dermisek:2005ij,Dermisek:2006dc} which is based on an \SO{10} GUT with a $D_3 \times [\U{1} \times \mathbb{Z}_2 \times \mathbb{Z}_3]$ family symmetry.   This model was shown
to give good fits to precision electroweak data, including quark, charged lepton and neutrino masses and mixing angles (see most recently the global $\chi^2$ analysis in  \cite{Albrecht:2007ii}).   In light of the observation of $\sin^2 \theta_{13}$ it is again a perfect time to re-analyze this model.  We are also able to compare the third family Yukawa unification analysis with the three family analysis which now includes hierarchical Yukawa matrices with unification of the (3,3) element of the Yukawa matrices.   Hence, off-diagonal elements in the Yukawa matrices give small corrections to exact Yukawa unification.

The paper is organized as follows. In \ref{model}, we present the \SO{10} model. In \ref{procedure}, we present the procedure used in the paper for analyzing the model. In \ref{3rd family}, we consider a model with gauge coupling unification and only the Yukawa couplings for the third family, which are assumed to unify at the GUT scale.   We perform a global $\chi^2$ analysis fitting the relevant low energy observables.   In \ref{three family}, we extend the analysis to all three families of quarks and leptons using a particular \SO{10} GUT model.   In this case,  we look for the minimum values of $\chi^2$ for five different choices of the universal squark and slepton mass,  $m_{16}$, defined at the GUT scale, $M_{G}$.  Finally,  the summary and conclusions are given in \ref{conclusion}.

\section{The Model \label{model}}
\subsection{Third family model}

Fermion masses and quark mixing angles are manifestly hierarchical.   The simplest way to describe this hierarchy is with Yukawa matrices which
are also hierarchical.   Moreover the most natural way to obtain the hierarchy is in terms of effective higher dimension operators of the form
\begin{equation}  W \supset \lambda \ 16_3 \ 10 \ 16_3 + 16_3 \ 10 \ \frac{45}{M} \ 16_2 + \cdots .
\end{equation}
This version of \SO{10} models has the nice features that it only requires small representations of \SO{10},  has many predictions
and can, in principle, find an UV completion in string theory. The only renormalizable term in $W$ is $\lambda \ 16_3 \ 10 \ 16_3$ which gives Yukawa coupling unification
\begin{equation}  \lambda = \lambda_t = \lambda_b = \lambda_\tau = \lambda_{\nu_\tau}  \end{equation} at $M_{GUT}$.
Note,  one {\it cannot} predict the top mass due to large SUSY threshold corrections to the bottom and tau masses, as shown in
\cite{Hall:1993gn,Carena:1994bv,Blazek:1995nv}.  These corrections are of the form
\begin{equation}  \delta m_b/m_b  \propto \frac{\alpha_3 \ \mu \ M_{\tilde g} \ \tan\beta}{m_{\tilde b}^2} +
\frac{\lambda_t^2 \ \mu \ A_t \ \tan\beta}{m_{\tilde t}^2} + {\rm log \ corrections} .
\end{equation} So instead  we use  Yukawa unification to predict the soft SUSY breaking masses. In order to fit the data,
we need \begin{equation} \delta m_b/m_b \sim - 2\% . \end{equation}  We take $\mu$, $M_{\tilde g} > 0$, thus
we need $\mu$, $A_t < 0$.  For a short list of references on this subject, see \cite{Blazek:2001sb,Blazek:2002ta,Baer:2001yy,Auto:2003ys,Tobe:2003bc,Dermisek:2003vn,Dermisek:2005sw,Baer:2008jn,Baer:2008xc}.

Given the following GUT scale boundary conditions, namely universal squark and slepton masses,  $m_{16}$,  universal cubic scalar parameter, $A_0$,
universal gaugino masses, $M_{1/2}$, and non-universal Higgs masses [NUHM] or ``just so'' Higgs splitting,  $m_{H_u}, \ m_{H_d}$  or
$m_{H_{u (d)}}^2 = m_{10}^2 [ 1 - (+) \Delta_{m_H}^2 ]$, we find that fitting the top, bottom and tau mass forces us into the region of SUSY breaking parameter
space with
\begin{equation}   A_0 \approx  - 2 m_{16},   \;\;  m_{10} \approx  \sqrt{2} \ m_{16}, \;\;  m_{16} > \ {\rm few \ TeV}, \;\;  \mu, M_{1/2} \ll m_{16}; \label{scrunch} \end{equation}
and, finally,  \begin{equation} \tan\beta \approx 50 . \end{equation}  In addition, radiative electroweak symmetry breaking requires
$\Delta_{m_H}^2 \approx 13\%$, with roughly half of this coming naturally from the renormalization group running of neutrino Yukawa couplings
from $M_G$ to $M_{N_\tau} \sim 10^{13}$ GeV \cite{Blazek:2002ta}.

It is very interesting that the above region in SUSY parameter space results in an inverted scalar mass hierarchy at the weak scale with the third family scalars
significantly lighter than the first two families \cite{Bagger:1999sy}.  This has the nice property of suppressing flavor changing
neutral current and CP violating processes.  These results depend solely on \SO{10} Yukawa unification for the third family.  In order to
demonstrate this, we perform a separate analysis with only third family observables (\ref{3rd family}) and then a complete three family analysis (\ref{three family}).\footnote{The large Yukawa coupling for the third family is the driving force for the inverted scalar mass hierarchy.   However, the particular boundary conditions of \ref{scrunch} were shown to maximize the effect.}

\subsection{Full Three Family Model \label{3family}}
\begin{figure}[h]
\centering
\subfigure[\footnotesize Renormalizable mass term for third family that gives rise to the $(3,3)$ element of the Yukawa matrix.]{
\includegraphics[width=0.45\textwidth]{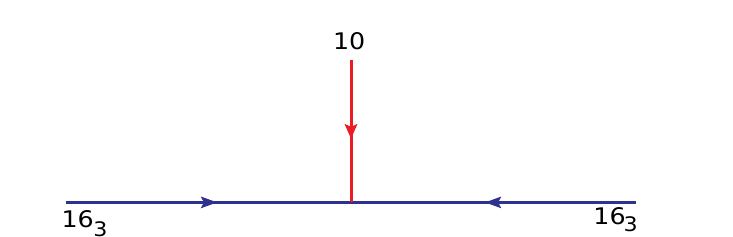}
}
\subfigure[\footnotesize Effective operator that generates the $(2,2)$ element of the Yukawa matrix.]{
\includegraphics[width=0.45\textwidth]{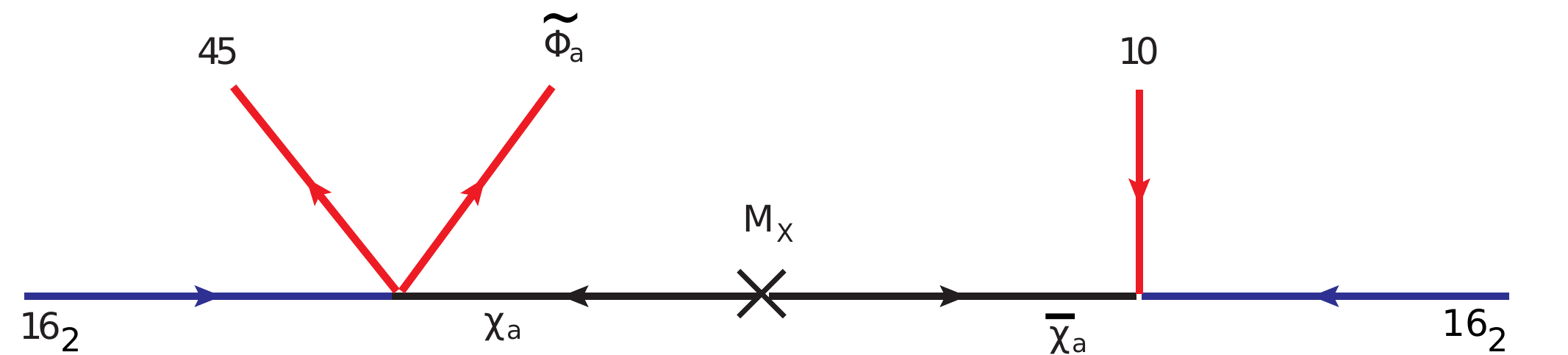}
}
\newline
\subfigure[\footnotesize Effective operators that generate the off-diagonal Yukawa couplings: $(b,c) = (3,2)$, $(2,3)$, $(3,1)$ or $(1,3)$]{
\includegraphics[width=0.45\textwidth]{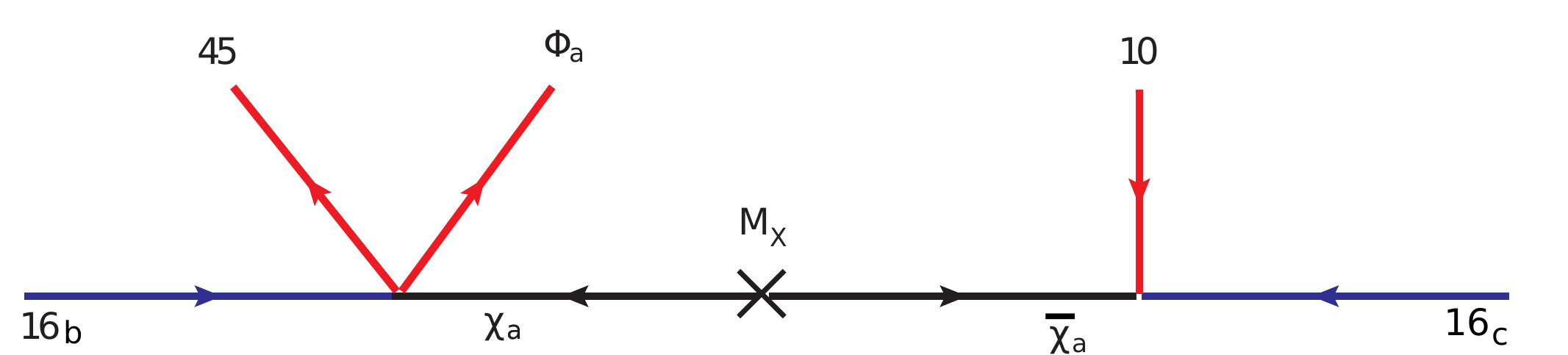}
}
\subfigure[\footnotesize  Effective operators that generate the off-diagonal Yukawa couplings: $(b,c) = (2,1)$ or $(1,2)$.]{
\includegraphics[width=0.45\textwidth]{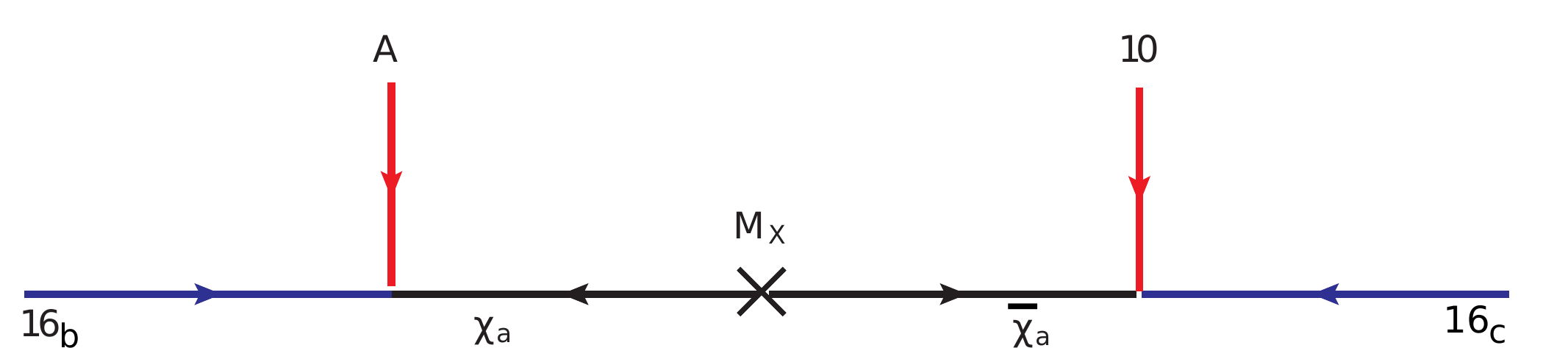}
}
\caption{\footnotesize The effective fermion mass operators obtained after integrating out the Froggatt-Nielsen massive states. Here, $a$ runs from 1 to 2.}
\label{massoperators}
\end{figure}

We now consider a complete three family \SO{10} model for fermion masses
and mixing, including neutrinos \cite{Dermisek:2005ij,Dermisek:2006dc,Albrecht:2007ii}.   The model also includes a
$D_3 \times [\U{1} \times \mathbb{Z}_2 \times \mathbb{Z}_3]$ family symmetry which is necessary to
obtain a predictive theory of fermion masses by reducing the number of arbitrary parameters in the Yukawa matrices.
Consider the superpotential generating the effective fermion Yukawa couplings:
\begin{equation} W_{\mathrm{ch.~fermions}} = \lambda \ 16_3 \ 10 \ 16_3 +  16_a \ 10 \ \chi_a +  \bar \chi_a \ ( M_{\chi} \ \chi_a + \ 45 \ \frac{\phi_a}{\hat M} \ 16_3 \ + \ 45 \ \frac{\tilde \phi_a}{\hat M} \  16_a + {\bs{A}} \ 16_a )
\label{Wchf}
\end{equation}
where \bs{45} is an \SO{10} adjoint field which is assumed to obtain
a VEV in the B -- L direction, $M_\chi$ is a linear combination of an
\SO{10} singlet and adjoint, and the index $a = 1,2$. Its VEV $M_\chi = M_0 ( 1 + \a X + \b Y)$ gives mass
to Froggatt-Nielsen states \cite{Froggatt:1978nt}. Here, $X$ and $Y$ are elements of the Lie
algebra of \SO{10} with $X$ in the direction of the \U{1} which
commutes with \SU{5} and $Y$ the standard weak hypercharge, and $ \a $ ,
$ \b $ are arbitrary constants which are fit to the data.  $\hat M$ is an \SO{10} invariant mass scale which in principle could
be obtained by integrating out additional Froggatt-Nielsen states.  Note that both $M_0$ and $\hat M$ are assumed to be above the
GUT scale. $\p$, $\ptil$, $\bs{A}$ are \SO{10} singlet ``flavon'' fields, $\bs{A}$ is a non-trivial one dimensional representation under $D_3$, and $\bar \chi_a$, $\chi_a$ are a pair of Froggatt-Nielsen states transforming as a \bsb{16} and \bs{16} under \SO{10}.   The so-called flavon fields are {\em assumed} to obtain VEVs of the form
\begin{equation}  \langle \p \rangle = \left( \begin{array}{c} \phi_1 \\ \phi_2 \end{array} \right), \;\;
\langle \ptil \rangle = \left( \begin{array}{c} 0 \\ \tilde \phi_2 \end{array} \right). \end{equation}
After integrating out the Froggatt-Nielsen states one obtains the effective fermion mass operators in \ref{massoperators}.

Inserting the flavon VEVs, one then obtains Yukawa matrices for
up-quarks, down-quarks, charged leptons and neutrinos given by
\begin{eqnarray}
Y_u =&  \left(\begin{array}{ccc}  0  & \epsilon' \ \rho & - \epsilon \ \xi  \\
             - \epsilon' \ \rho &  \tilde \epsilon \ \rho & - \epsilon     \\
       \epsilon \ \xi   & \epsilon & 1 \end{array} \right) \; \lambda & \nonumber \\
Y_d =&  \left(\begin{array}{ccc}  0 & \epsilon'  & - \epsilon \ \xi \ \sigma \\
- \epsilon'   &  \tilde \epsilon  & - \epsilon \ \sigma \\
\epsilon \ \xi  & \epsilon & 1 \end{array} \right) \; \lambda & \label{eq:yukawaD31} \\
Y_e =&  \left(\begin{array}{ccc}  0  & - \epsilon'  & 3 \ \epsilon \ \xi \\
          \epsilon'  &  3 \ \tilde \epsilon  & 3 \ \epsilon  \\
 - 3 \ \epsilon \ \xi \ \sigma  & - 3 \ \epsilon \ \sigma & 1 \end{array} \right) \; \lambda &
 \nonumber \end{eqnarray}
with  \begin{eqnarray}  \xi \;\; =  \;\; \phi_2/\phi_1, & \;\;
\tilde \epsilon  \;\; \propto   \;\; \tilde \phi_2/\hat M,  & \label{eq:omegaD3} \\
\epsilon \;\; \propto  \;\; \phi_1/\hat M, &  \;\;
\epsilon^\prime \;\; \sim  \;\;  ({\bf A}/M_0), \nonumber \\
  \sigma \;\; =   \;\; \frac{1+\alpha}{1-3\alpha}, &  \;\; \rho \;\; \sim   \;\;
  \beta \ll \alpha &  \nonumber \eea and
\bea Y_{\nu} =&  \left(\begin{array}{ccc}  0  & - \epsilon' \ \omega & {3 \over 2} \ \epsilon \ \xi \ \omega \\
      \epsilon'  \ \omega &  3 \ \tilde \epsilon \  \omega & {3 \over 2} \ \epsilon \ \omega \\
       - 3 \ \epsilon \ \xi \ \sigma   & - 3 \ \epsilon \ \sigma & 1 \end{array} \right) \; \lambda & \label{eq:yukawaD32}
  \end{eqnarray}  with $\omega \;\; =  \;\; 2 \, \sigma/( 2 \, \sigma - 1)$ and a Dirac neutrino mass matrix given by
 \begin{equation} m_\nu \equiv Y_\nu \frac{v}{\sqrt{2}} \sin\beta.
 \label{eq:mnuD3}
  \end{equation}
From \ref{eq:yukawaD31} and \ref{eq:yukawaD32} one can see that the flavor hierarchies in the
Yukawa couplings are encoded in terms of the four complex parameters
$\rho, \sigma, \tilde \varepsilon, \xi$ and the additional real ones
$\varepsilon, \varepsilon', \lambda$.
These matrices contain 7 real parameters and 4 arbitrary phases.
Note,  the superpotential (\ref{Wchf}) has many arbitrary parameters.   However, at the end of the day the effective Yukawa matrices have much fewer parameters.
This is good, because we then obtain a very predictive theory.   Also, the quark mass matrices accommodate the Georgi-Jarlskog mechanism, such that
$m_\mu/m_e \approx 9 m_s/m_d$.  This is a result of the \bs{45} VEV in the $B - L$ direction.

We then add 3 real Majorana mass parameters for the neutrino see-saw mechanism. The anti-neutrinos obtain GUT scale masses by mixing with
three \SO{10} singlets ($N_a$ for $a = 1,2$ and $N_3$) transforming as a $D_3$ doublet and singlet, respectively. The full superpotential is given by $W =
W_{\mathrm{ch.~fermions}} + W_{\mathrm{neutrino}}$ with
\begin{equation}
W_{\mathrm{neutrino}} = \overline{16} \left(\lambda_2 \ N_a \ 16_a \ + \ \lambda_3 \ N_3 \ 16_3 \right) +  \frac{1}{2} \left(S_{a} \ N_a \ N_a + S_3 \ N_3 \ N_3\right)
\label{eq:WneutrinoD3}
\end{equation}
where the fields $S_a, \ S_3$ are additional flavon fields whose VEVs provide Majorana masses for the states $N_a, \ N_3$.
We assume $\overline{16}$ obtains a VEV, $v_{16}$, in the right-handed neutrino direction, and $\langle S_{a}
\rangle = M_a$ for $a = 1,2$ and $\langle S_3 \rangle = M_3$.  The effective neutrino mass terms are given by
\begin{equation} W =  \nu \ m_\nu \ \bar \nu + \bar \nu \ V \ N + \frac{1}{2} \ N \ M_N \ N \end{equation} with
\begin{equation} V = v_{16} \ \left(
\begin{array}{ccc} 0 &  \lambda_2 & 0 \\
\lambda_2 & 0 & 0 \\ 0 & 0 &  \lambda_3 \end{array} \right), \; M_N = {\rm diag} ( M_1,\ M_2,\ M_3)  \end{equation}  all assumed
to be real.  Finally, upon integrating out the heavy Majorana neutrinos we obtain the $3 \times 3$ Majorana mass matrix for the light
neutrinos in the lepton flavor basis given by
\begin{equation}
{\cal M} =   U_e^T \ m_\nu  \ M_R^{-1} \ m_\nu^T  \ U_e ,
\end{equation}
where the effective right-handed neutrino Majorana mass matrix is given by:
\begin{equation}
M_R =  V \ M_N^{-1}  \ V^T  \  \equiv \  {\rm diag} ( M_{R_1}, M_{R_2}, M_{R_3} ),
\end{equation}
with \begin{eqnarray} M_{R_1} = (\lambda_2 \ v_{16})^2/M_2, \quad  M_{R_2} = (\lambda_2 \ v_{16})^2/M_1, \quad  M_{R_3} =
(\lambda_3 \ v_{16})^2/M_3 . \label{eq:rhmass} \end{eqnarray}

\section{Procedure \label{procedure}}

\begin{table}
\begin{center}
\renewcommand{\arraystretch}{1.2}
\scalebox{0.83}{
\begin{tabular}{|l||c|c||c|c||}
\hline
Sector &  Third Family Analysis & \# & Full three family Analysis & \# \\
\hline
gauge             & $\alpha_G$, $M_G$, $\epsilon_3$                   & 3  & $\alpha_G$, $M_G$, $\epsilon_3$                                                 & 3 \\
SUSY (GUT scale)  & $m_{16}$, $M_{1/2}$, $A_0$, $m_{H_u}$, $m_{H_d}$  & 5  & $m_{16}$, $M_{1/2}$, $A_0$, $m_{H_u}$, $m_{H_d}$                                & 5 \\
textures          & $\lambda$                                         & 1  & $\epsilon$, $\epsilon'$, $\lambda$, $\rho$, $\sigma$, $\tilde \epsilon$, $\xi$  & 11\\
neutrino          &                                                   & 0  & $M_{R_1}$, $M_{R_2}$, $M_{R_3}$                                                 & 3 \\
SUSY (EW scale)   & $\tan \beta$, $\mu$                               & 2  & $\tan \beta$, $\mu$                                                             & 2 \\
\hline
Total \#          &                                                   & 11 &                                                                                 & 24\\
\hline
\end{tabular}
}
\caption{\footnotesize The model is defined by three gauge parameters, $\alpha_{G}, M_{G}, \epsilon_3$; one large Yukawa coupling, $\lambda$; 5 SUSY parameters defined at the GUT scale, $m_{16}$ (universal scalar mass for squarks and sleptons), $M_{1/2}$ (universal gaugino mass), $m_{H_u}, \ m_{H_d}$ (up and down Higgs masses), and $A_0$ (universal trilinear scalar coupling); $\mu, \ \tan\beta$ obtained at the weak scale by consistent electroweak symmetry breaking. The full three family model has additional off-diagonal Yukawa couplings, and includes 3 right-handed neutrino masses.}
\label{tab:parameters}
\end{center}
\end{table}

\subsection*{Renormalization Group Equations}

The model parameters, summarized in \ref{tab:parameters}, are defined at the grand unification scale $M_G$ with the exception of $\tan\beta$ and $\mu$ that are defined at the electroweak scale. At the GUT scale, $\alpha_G \equiv \alpha_1(M_G) = \alpha_2(M_G)$ and $\alpha_3(M_G) = \alpha_G (1 +\epsilon_3)$, where $\epsilon_3$ is the GUT scale threshold correction\footnote{Without presenting a complete GUT we leave $\epsilon_3$ as a free parameter.
In this way, our analysis will also apply to orbifold GUTs or string compactifications with a scale of order $M_{G}$.} necessary to fit the strong coupling to experimental data at the electroweak scale, $M_Z$. These 3 gauge parameters, the 11 Yukawa textures (described in \ref{3family}), 5 SUSY boundary conditions, and 3 real neutrino mass parameters allow us to completely define the model at the GUT scale and derive all the parameters of the  minimal supersymmetric standard model (MSSM).

\begin{figure}[p]
\thisfloatpagestyle{empty}
\centering
\includegraphics[width=1.0\textwidth]{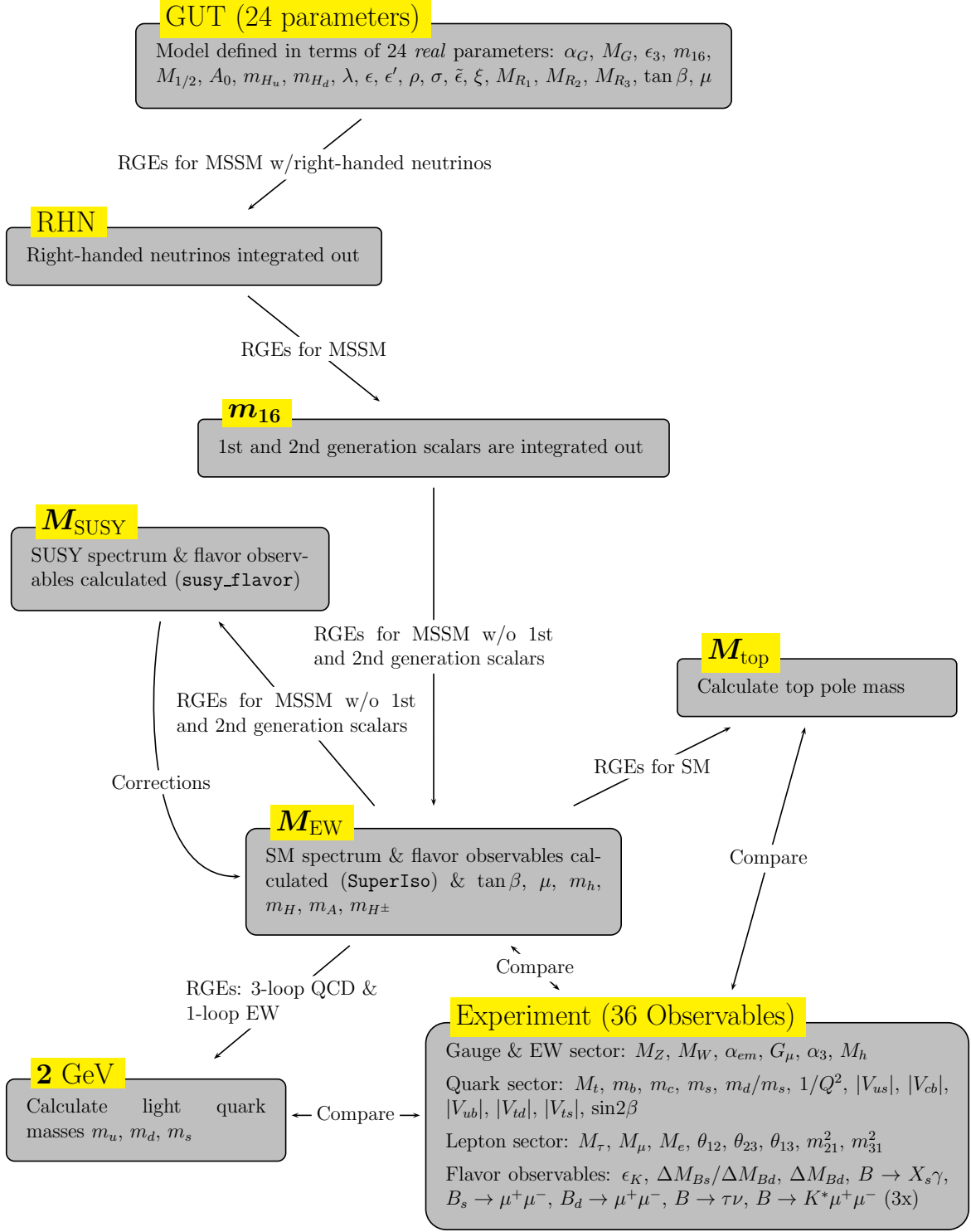}
\caption{\footnotesize{This schematic shows the steps that must be employed to evolve a GUT model to the low energies and calculate observables at the relevant scales to compare with experimental data. Note that we use threshold corrections instead of integrating out two family of scalars from the particle spectrum.
}}
\label{fig:flowchart}
\end{figure}

First, the GUT scale parameters are RGE evolved to the right-handed neutrino scale where the RH neutrinos are integrated out (see \vref{fig:flowchart}). The right-handed neutrinos have three different scales associated with them, and the most relevant one is the third-family RHN that is mostly responsible for splitting the up and down type Higgs masses. We therefore choose to integrate out all the right-handed neutrinos at one single scale, $M_{N_\tau} = M_{R_3}$.

Below the scale of the RHNs, we use the 2-loop MSSM RGEs for both dimensionful and dimensionless parameters. Ideally, one should evolve all parameters to the scale of the heavy scalars ($m_{16}$ in this case, as shown in \ref{fig:flowchart}) and integrate them out and proceed to evolve to the weak scale using an effective theory without the first two generation scalars. We choose an alternative approach and use the 2-loop MSSM RGE\footnote{In scenarios with heavy scalars, it has been shown that the 2-loop contributions to the third generation scalars can lead to dramatic consequences, like driving the stop mass squared negative \cite{ArkaniHamed:1997ab} and thus it is important to include the 2-loop RGEs in scenarios such as discussed here.} evolution down to the weak scale and correct for the additional running by including 1-loop threshold corrections to the relevant observables\footnote{For the calculation of Higgs mass, we define an effective theory at the scale $M_{\mathrm{SUSY}}$ and interface our calculation with the code by authors in Ref.\cite{Bernal:2007uv}}. This approximation eliminates the need to define multiple effective theories. In our analysis, we have been careful to take into account the corresponding threshold corrections for all observables.

\begin{table}[p!]
\begin{center}
\scalebox{0.8}{
\renewcommand{\arraystretch}{1.2}
\begin{tabular}{|l|l|l|l|c|}
\hline
\textbf{Observable} &  \textbf{Exp.~Value}   & \textbf{Ref.}          & \textbf{Program} &  \textbf{Th.~Error}  \\
\hline
\hline
$\alpha_3(M_Z)$          &  $0.1184\pm0.0007$               & \cite{Beringer:1900zz} & \maton{}    &  $0.5\%$     \\
$\alpha_\text{em}$  &  $1/137.035999074(44)$            & \cite{Beringer:1900zz} & \maton{}    &  $0.5\%$     \\
$G_\mu$             &  $1.16637876(7)\times10^{-5}\text{ GeV}^{-2}$ & \cite{Beringer:1900zz} & \maton{}    &  $1\%$     \\
$M_W$               &  $80.385\pm0.015\text{ GeV}$                     & \cite{Beringer:1900zz} & \maton{}    &  $0.5\%$     \\
$M_Z$              &   $91.1876 \pm 0.0021$                             &   \cite{Beringer:1900zz}        &  Input          &     $0.0\%$   \\
\hline
$M_t$               &  $173.5\pm1.0\text{ GeV}$                & \cite{Beringer:1900zz} & \maton{}    &  $0.5\%$     \\
$m_b(m_b)$               &  $4.18\pm0.03\text{ GeV}$ & \cite{Beringer:1900zz} & \maton{}    &  $0.5\%$     \\
$m_c(m_c)$               &  $1.275\pm0.025\text{ GeV}$                 & \cite{Beringer:1900zz} & \maton{}    &  $0.5\%$     \\
$m_s(2\text{ GeV})$               &  $95\pm5\text{ MeV}$                & \cite{Beringer:1900zz} & \maton{}    &  $0.5\%$     \\
$m_s/m_d\,(2\text{ GeV})$          &  $17-22$               & \cite{Beringer:1900zz} & \maton{}    &  $0.5\%$     \\
$Q$                 &  $21-25$              & \cite{Beringer:1900zz} & \maton{}    &  $5\%$    \\
$|V_{us}|$            &  $0.2252\pm0.0009$               & \cite{Beringer:1900zz} & \maton{}    &  $0.5\%$     \\
$|V_{ub}|$            &  $0.00377\pm0.00085$              & \cite{Beringer:1900zz} & \maton{}    &  $0.5\%$     \\
$|V_{cb}|$            &  $0.04065\pm0.00195$               & \cite{Beringer:1900zz} & \maton{}    &  $0.5\%$     \\
$|V_{td}|$            &  $0.00840\pm0.0006$               & \cite{Beringer:1900zz} & \maton{}    &  $0.5\%$     \\
$|V_{ts}|$            &  $0.0429\pm0.0026$               & \cite{Beringer:1900zz} & \maton{}    &  $0.5\%$     \\
$\sin2\beta$        &  $0.679\pm0.020$                & \cite{Beringer:1900zz} & \maton{}    &  $0.5\%$     \\
$M_\tau$            &  $1776.82\pm0.16\text{ MeV}$                & \cite{Beringer:1900zz} & \maton{}    &  $0.5\%$     \\
$M_\mu$             &  $105.6583715(35)\text{ MeV}$       & \cite{Beringer:1900zz} & \maton{}    &  $0.5\%$     \\
$M_e$               &  $0.510998928(11)\text{ MeV}$     & \cite{Beringer:1900zz} & \maton{}    &  $0.5\%$     \\
\hline
$M_h$               &  $125.3\pm0.4\pm0.5\text{ GeV}$                & \cite{:2012gu} & Ref.~\cite{Bernal:2007uv}  &  $3$ GeV     \\
\hline
$\sin^2\theta_{12}$ &  $0.27-0.34$ (3$\sigma$ range)                & \cite{NuFIT:2012-10-24} & \maton{}    &  $0.5\%$     \\
$\sin^2\theta_{23}$ &  $0.34-0.67$ (3$\sigma$ range)                & \cite{NuFIT:2012-10-24} & \maton{}    &  $0.5\%$     \\
$\sin^2\theta_{13}$ &  $0.016-0.030$ (3$\sigma$ range)               & \cite{NuFIT:2012-10-24} & \maton{}    &  $0.5\%$     \\
$\Delta m_{21}^2$   &  $(7.00-8.09)\times10^{-5}\text{ eV}^2$ (3$\sigma$ range)      & \cite{NuFIT:2012-10-24} & \maton{}    &  $0.5\%$     \\
$\Delta m_{31}^2$   &  $(2.27-2.69)\times10^{-3}\text{ eV}^2$ (3$\sigma$ range)      & \cite{NuFIT:2012-10-24} & \maton{}    &  $0.5\%$     \\
\hline
$\text{BR}(b \rightarrow s \gamma)$                &  $(343\pm21\pm7) \times 10^{-6}$      & \cite{hfag:2012-10-24} & \superiso{}      &  $(181 - 505) \times 10^{-6}$    \\
$\text{BR}(B \rightarrow K^* \mu \mu)_{1\leq q^2\leq 6\text{ GeV}^2}$  &  $(1.97\pm0.21)\times10^{-7}$      & \cite{hfag:2012-10-24} & \superiso{}      &  $(0.79-3.15) \times 10^{-7}$  \\
$\text{BR}(B \rightarrow K^* \mu \mu)_{14.18\leq q^2\leq 16\text{ GeV}^2}$ &  $1.20^{+0.11}_{-0.10}\times10^{-7}$      & \cite{hfag:2012-10-24} & \superiso{}      &  $(0.48 - 1.92)\times 10^{-7}$  \\
$q_0^2(\text{A}_\text{FB}(B \rightarrow K^* \mu \mu))$     &  $4.9^{+1.1}_{-1.3}\text{ GeV}^2$      & \cite{LHCb-CONF-2012-008} & \superiso{}      &  $4.86 - 4.94$  \\
$\text{BR}(B_s  \rightarrow \mu^+ \mu^-) $          &  $3.2\times 10^{-9}$      & \cite{:2012ct} & \susyflavor{}    &  $ 1.5 \times 10^{-9}$     \\
$\text{BR}(B_u  \rightarrow \tau \nu) $             &  $(166\pm33)\times 10^{-6}$      & \cite{hfag:2012-10-24} & \susyflavor{}    &  $ (83 - 249)\times 10^{-6}$  \\
$\text{BR}(B_d  \rightarrow \mu^+ \mu^-) $          &  $<8.1\times 10^{-10}$      & \cite{hfag:2012-10-24} & \susyflavor{}    &  $ < 9.72 \times 10^{-10}$  \\
$\Delta m_{B_d}$                                 &  $(3.337\pm0.033)\times 10^{-10}\text{ MeV}$      & \cite{Beringer:1900zz} & \susyflavor{}    &  $(2.67 - 4.00)\times 10^{-10}$    \\
$\Delta m_{B_s} / \Delta m_{B_d}$                 &  $35.06\pm0.42$     & \cite{Beringer:1900zz} & \susyflavor{}    &  $28.05 - 42.07$     \\
$\epsilon_K$                               &  $(2.228\pm0.11)\times 10^{-3}$      & \cite{Beringer:1900zz} & \susyflavor{}    &  $(2.00 - 2.45)\times 10^{-3} $     \\
\hline
\end{tabular}
}
\end{center}
\caption{\footnotesize The 36 observables that we fit and their experimental values. In the 4th column, we indicate the software package that gives us the theoretical prediction. In the last column, we show what we have assumed for the theoretical errors. Here, $Q^2=(m_s^2-1/4(m_u+m_d)^2)/(m_d^2-m_u^2)$ is defined on p.~657 of Ref.~\cite{Beringer:1900zz}. The number(s) in brackets after some of the values indicate the $1\sigma$ uncertainty in the last digit(s). Capital letters denote pole masses. We take LHCb results into account, but use the average by Ref.~\cite{hfag:2012-10-24}. All experimental errors are $1\sigma$ unless otherwise indicated.  To account for the inconsistencies in the inclusive and exclusive measurements of $|V_{ub}|$ and $|V_{cb}|$, we allow our result to be within the experimental error from both the inclusive and the exclusive measurement. To minimize theoretical uncertainties, we fit the ratio $\Delta m_{B_s} / \Delta m_{B_d}$ and derive its error by the usual formula for error propagation using the value $\Delta m_{B_s}=(117.0\pm0.8)\times10^{-10}\text{ MeV}$ \cite{Beringer:1900zz} and assuming no correlations between the errors. Finally, the $Z$ mass is
fit precisely via a separate $\chi^2$ function solely imposing electroweak symmetry breaking.
}
\label{tab:allobservables}
\end{table}

\subsection*{Electroweak Observables}

 At the weak scale, we calculate the SUSY spectrum and the SUSY threshold corrections to the fermion masses and CKM matrix elements. Especially in the large $\tan\beta$ regime, these SUSY threshold corrections are very important for the down type quarks and charged leptons and can be at the percent level in Yukawa-unified SUSY models \cite{Blazek:1995nv}. We then use the threshold corrected fermion masses to determine the tree level masses for the squarks and sleptons. In addition, we also determine the one-loop pole mass for gluino and the CP-odd Higgs mass.  The precision electroweak observables $M_Z$, $M_W$, $G_{\mu}$, $\alpha_{em}^{-1}(M_Z)$, $\alpha_s(M_Z)$ are calculated including 1-loop threshold corrections, using the procedure described in \cite{Pierce:1996zz, Chankowski:1994ua}. Following the prescription in \cite{Pierce:1996zz}, the condition for consistent radiative electroweak symmetry breaking is also imposed at the weak scale, and for this, we use the physical $Z$ pole mass. The parameter $\mu$ is fixed by this procedure via a separate $\chi^2$ minimization, and in the process, we fit the $Z$ mass precisely to the physical $Z$ pole mass. In the calculation of $M_Z$ and $M_W$, we only include the 1-loop corrections from the third family scalars, since the first two generation scalars are integrated out at $m_{16}$.  We assign a theoretical uncertainty of 0.5\% to our calculation of the electroweak observables (except for $M_Z$) due to the approximate treatment of thresholds described above. We also assign a 1\% theoretical uncertainty to our calculation of $G_{\mu}$, since we neglect the SUSY vertex and box diagrams. Finally, to compare to experiment, $\alpha_{em}$ is evolved to zero momentum transfer.

\subsection*{Charged Fermion masses and mixing angles}

Below $M_Z$, we integrate out all SUSY partners and electroweak gauge bosons to obtain an effective $\SU{3}\times\U{1}_{\mathrm{em}}$ low energy theory. We use 1-loop QED and 3-loop QCD RGEs to renormalize to the appropriate scales and calculate the low energy observables. We fit the top quark pole mass, and the bottom and charm quark $\overline{\mathrm{MS}}$ masses are calculated at their respective masses. All the other light quark masses are calculated at the scale of 2 GeV. We fit 7 observables relevant to quark masses, 3 charged lepton masses, and 6 CKM observables. The theoretical uncertainty in their calculation is again estimated to be 0.5 \%. Since the light quark masses are not measured to very high precision, we choose to fit multiple correlated observables. These include the $\overline{\mathrm{MS}}$ strange quark mass, the mass ratio $m_d/m_s$ and the mass ratio Q defined in the PDG \cite{Beringer:1900zz} as
\begin{equation}
Q^2=\frac{m_s^2-1/4(m_u+m_d)^2}{m_d^2-m_u^2}, \quad \text{or equivalently,} \quad \left(\frac{m_u}{m_d}\right)^2 + \frac{1}{Q^2} \left(\frac{m_s}{m_d}\right)^2 = 1
\label{ellipse}
\end{equation}
The CKM matrix is calculated from the left and right mixing matrices by diagonalizing the Yukawa matrices and including the SUSY threshold corrections. 6 CKM observables ($|V_{us}|$, $|V_{ub}|$, $|V_{cb}|$, $|V_{td}|$, $|V_{ts}|$ and $\sin2\beta$) are included in our global fit analysis. To account for the inconsistencies in the inclusive and exclusive measurements of $|V_{ub}|$ and $|V_{cb}|$, we allow our result to be within the experimental error from both the inclusive and the exclusive measurement. The pole masses in the lepton sector are calculated with 1-loop electromagnetic threshold corrections.

To execute the steps elaborated so far, we use a code \maton{}\label{def:maton}, originally developed by Radovan Derm\'{i}\v{s}ek to study Yukawa unification in the \SO{10} model with $D_3 \times [\U{1} \times \mathbb{Z}_2 \times \mathbb{Z}_3]$ family symmetry \cite{Dermisek:2006dc}. \maton{} has been restructured and extended appropriately to adapt to the current analysis.

\subsection*{Higgs Mass}

 The recent observation of the Higgs boson at the LHC \cite{:2012gk,:2012gu} will allow us to highly constrain the parameter space of the model. Flavor constraints have already pushed the first two generation scalars of Yukawa-unified SUSY models $\gtrsim$ 10 TeV \cite{Altmannshofer:2008vr}. In contrast, the third family scalars have mass about a few TeV, purely by the effects of RGE running. The hierarchy between the first two and the third generations alleviates the constraints from flavor physics and CP violating observables, and at the same time eases the large fine-tuning in models with heavy scalars. In addition to the TeV range scalars, the large $A$-terms make it easy to obtain a Higgs mass of about 125 GeV. We integrate out all the scalars (including the third generation squarks and sleptons) below the scale $M_{\mathrm{SUSY}}$, and calculate the Higgs boson mass using the dedicated  code by the authors of Ref. \cite{Bernal:2007uv}, that is best suited to our case where the sfermions are very heavy. Given the boundary conditions,
\begin{equation}
\mu(M_Z),\ M_1(M_{\mathrm{SUSY}}),\ M_2(M_{\mathrm{SUSY}}),\ M_3(M_{\mathrm{SUSY}}),\ M_{\mathrm{SUSY}},\ \tan\beta,\ A_t(M_{\mathrm{SUSY}}) \nonumber
\end{equation}
at the scale $M_{\mathrm{SUSY}} = \sqrt{M_{\tilde{t}_1} \times M_{\tilde{t}_2}}$, (where $M_1,\ M_2,\ M_3$ are the gaugino masses at the scale $M_{\mathrm{SUSY}}$), the routine \cite{Bernal:2007uv} determines the Higgs mass by calculating the corrections to the Higgs quartic coupling:
\begin{equation}
 M_h = \sqrt{\frac{\lambda(M_{\mathrm{SUSY}})}{\sqrt{2} G_\mu}} \left( 1  + \delta^{SM} (M_{\mathrm{SUSY}}) + \delta^{\chi} (M_{\mathrm{SUSY}}) \right)
\end{equation}
$\delta^{\chi}$ are the contributions from chargino and neutralino diagrams. The quartic coupling $\lambda(M_{\mathrm{SUSY}})$ is given by:
\begin{equation}
 \lambda(M_{\mathrm{SUSY}}) = \frac{1}{4} \left(g^2 + g'^2 \right) + \frac{3 h_t^4}{8 \pi^2} \left[ \left(1- \frac{g^2 + g'^2} {8 h_t^2} \right) \frac{X_t^2}{M_{\mathrm{SUSY}}^2} - \frac{X_t^4}{12 M_{\mathrm{SUSY}}^4}  \right]
\end{equation}

We have to point out an important difference in our approach. The conventional method is to use the SM inputs of $M_Z$, $G_{\mu}$, $\alpha_{em}^{-1} (M_Z)$, $\alpha_s(M_Z)$, $M_t$, $m_b(m_b)$, $M_{\tau}$ to determine the gauge and the Yukawa couplings at the scales $M_{\mathrm{SUSY}}$ and further constrain the GUT scale parameters. We instead like to predict these low energy observables and constrain the GUT scale parameter space based on a global $\chi^2$ fit to the data. In our calculation of the Higgs mass, we take the gauge and Yukawa couplings as input at the scale $M_{\mathrm{SUSY}}$, obtained from RGE evolution using \maton{} and calculate the Higgs mass using these inputs. The approach we adopt here is purely top-down. We have adapted the routine \cite{Bernal:2007uv} to suit this line of analysis. Nevertheless, we have compared the spectrum we obtain from \maton{} with that from \texttt{softsusy}\footnote{Without making significant changes to \texttt{softsusy} or other publicly available codes, we find that we can only make rough comparisons of the spectra. This is because to the best of our knowledge, most of the currently available codes do not handle complex parameters. In addition, many do not include right-handed neutrinos, and do not offer an easy way to implement the particular GUT scale Yukawa texture of the model.} \cite{Allanach:2001kg} and find good agreement.

\subsection*{Neutrino Sector}

We are fitting 5 observables in the neutrino sector: the mixing angles $\theta_{12}$, $\theta_{23}$, $\theta_{13}$, and the mass-squared differences $\Delta m_{31}\equiv m_3^2-m_1^2$ and $\Delta m_{21}\equiv m_2^2-m_1^2$ (cf.~\ref{tab:allobservables}). The most dramatic change in the experimental determination of the neutrino parameters in recent years comes from the Daya Bay and Reno collaborations \cite{An:2012eh,Ahn:2012nd} that have confirmed that $\theta_{13}\sim9^\circ$ is indeed large. Moreover, there are tentative hints that $\theta_{23}$ is not maximal \cite{Fogli:2012ua, GonzalezGarcia:2012sz}. Whereas Ref.~\cite{Fogli:2012ua} sees a preference at $\sim2\sigma-3\sigma$ for the first octant, i.e.~$\theta_{23}<45^\circ$, Ref.~\cite{GonzalezGarcia:2012sz} finds an equal probability for $\theta_{23}$ being larger or smaller than $45^\circ$. In the following, we will be using the best-fit values and the $3\sigma$ uncertainties quoted by the \nufit{} collaboration \cite{GonzalezGarcia:2012sz} which are in agreement with Ref.~\cite{Fogli:2012ua} at $3\sigma$.

\subsection*{Flavor Physics}

The strongest constraints on the model come from B-physics. For calculating the flavor observables, we use two publicly available codes, namely \susyflavor{} \cite{Crivellin:2012jv} and \superiso{} \cite{Mahmoudi:2007vz,Mahmoudi:2008tp}. Since the boundary conditions that we impose at the GUT scale may generate large off-diagonal and in general complex entries at the low scale, \susyflavor{} is better adapted to our needs. Note that \susyflavor{}, in contrast to comparable programs that calculate similar processes, does not assume minimal flavor violation (MFV), and allows for general, full three family, complex soft parameters. This is particularly important in our case, since we are calculating several CP violating observables and need to take into account\footnote{We calculate the particle spectrum using \maton{}, see comments on p.~\pageref{def:maton}. To the best of our knowledge, there is currently no publicly available spectrum generator that fully takes into account all the complex phases of the MSSM.} the complex phases in the soft parameters. Hence, \susyflavor{} is our default choice for all flavor observables with the following exceptions. For $B \rightarrow X_s \gamma$, we use \superiso{}, since \susyflavor{} does not include the NNLO SM corrections. We have verified that the discrepancy between \susyflavor{} and \superiso{} in the parameter space that is of interest to us is at most 10\% and typically less than 7\%. Also, we use \superiso{} for the observables connected to the decay process $B \rightarrow K^* \mu^+ \mu^-$, since \susyflavor{} does not provide them. It is important to note that \superiso{} has some built-in assumptions that prove to be too restrictive in our case. E.g.~\superiso{} assumes all soft parameters to be real, and only takes the diagonal entries of the third-family trilinear couplings into account. As a consequence, we have assigned larger theoretical uncertainties to the values calculated by \superiso{} (see \ref{tab:allobservables}). Additional sources of uncertainties in the flavor observables derive from the theoretical determination of the B meson decay constant and from the experimental measurements of the CKM matrix elements.

LHCb has recently measured \cite{:2012ct} the $\text{Br}(B_s \rightarrow \mu^+ \mu^-)$ which is in good agreement with the SM prediction. This pushes the CP-odd Higgs mass to a few TeV and hence leads to the Higgs decoupling limit.  {\em Thus the light Higgs is predicted to be SM-like.} The recent observation of zero-crossing in the forward-backward asymmetry of $B \rightarrow K^* \mu^+ \mu^- $ constrains the Wilson coefficient $C_7$ to be of the same sign as that in the SM. This imposes the additional constraint for the model that if $\mu>0$, in order to satisfy the branching fraction observed in the process $B \rightarrow X_s \gamma$ the first two generation scalars have to be heavier than at least 10 TeV.

\subsection*{Global Fit}

In the last step of our calculation, we construct a $\chi^2$ function in terms of the 36 calculated observables (see \ref{tab:allobservables}).
 \begin{equation}
\chi^2 = \sum_{i} \frac{|y_i - y_i^{data}|^2}{\sigma_i^2}
\end{equation}
 $y_i$ and $y_i^{data}$ are the theoretical prediction and experimental measurement, respectively, for each observable. $\sigma_i$ is the error on each observable, the theoretical and experimental errors added in quadrature. In the general case, we vary 23 parameters (see \ref{tab:parameters} and note that $m_{16}$ is fixed in all the analyses)  in order to fit 36 observables, which amounts to 12 (or 13 counting the separate fit to the Z pole mass) degrees of freedom (d.o.f.). We will consider the $\chi^2$ per d.o.f. for the model as a qualitative measure of the goodness of fit. We will look at the pulls from the individual observables to assess the goodness of fit of the model. 
 
 Finding the global minimum for a model with 23 parameters is a formidable task. In the present analysis, we minimize the $\chi^2$ function using the Minuit package maintained by CERN \cite{James:1975dr}. Note that Minuit is not guaranteed to find the \textit{global} minimum, but will in most cases converge on a local one. For that reason, we iterate $\mathcal{O}$(100) times the minimization procedure for each set of input parameters, and in each step we take a different initial guess for the minimum (required by Minuit) so that we have a fair chance of finding the true minimum. This, of course, requires large computing resources, and to that end we have used the Ohio Supercomputer Center in Columbus and the ``Centre de Calcul de l'Institut National de Physique Nucl\'{e}aire et Physique des Particules'' in Lyon.

\section{Third family analysis  \label{3rd family}}

In this section we analyze the consequences of Yukawa unification for the the
third family in the context of minimal \SO{10} supersymmetric grand unification
defined by the superpotential term, W $\supset \lambda \ 16_3 \ 10 \ 16_3$. The aim of this analysis is to study the SUSY spectrum, and we argue that the constraints on the SUSY spectrum come predominantly from
the third family, the lightest Higgs mass and the branching ratio $BR(B_s \rightarrow \mu^+ \ \mu^-)$.  There are
24 parameters in total in the DR model \cite{Dermisek:2005ij,Dermisek:2006dc}, and in this section we focus on 11
parameters (summarized in \ref{tab:parameters}) that are used to evaluate
11 low energy observables, $M_W$, $M_Z$, $G_F$, $\alpha_{em}^{-1}$,
$\alpha_s(M_Z)$, $M_t$, $m_b(m_b)$, $M_{\tau}$, $Br(B\rightarrow X_s \gamma)$,
$Br(B_s \rightarrow \mu^+ \mu^-)$, and the lightest Higgs mass, $M_h$. We specify the model with
the full 24 parameters, but we only vary 11 in the minimization procedure to fit
the 11 observables listed above.  The irrelevant parameters for this analysis,
namely, the neutrino parameters and the off-diagonal Yukawa textures, are set to
constant values and do not enter into the minimization procedure.\footnote{In this section, when calculating flavor violating
observables, we use \susyflavor{} with the experimental input values for the light fermion masses and mixing angles.} Similarly, the
low energy observables connected to the first two families do not enter the
$\chi^2$ function. The effects of the off-diagonal
Yukawa textures will be discussed in \ref{three family}.

\begin{figure}[h!]
\centering
\subfigure[\footnotesize With increasing $m_{16}$, $\chi^2$ first dramatically decreases, and after reaching a minimum around $m_{16}\simeq20$ TeV, starts increasing again.]{
\includegraphics[width=0.45\textwidth]{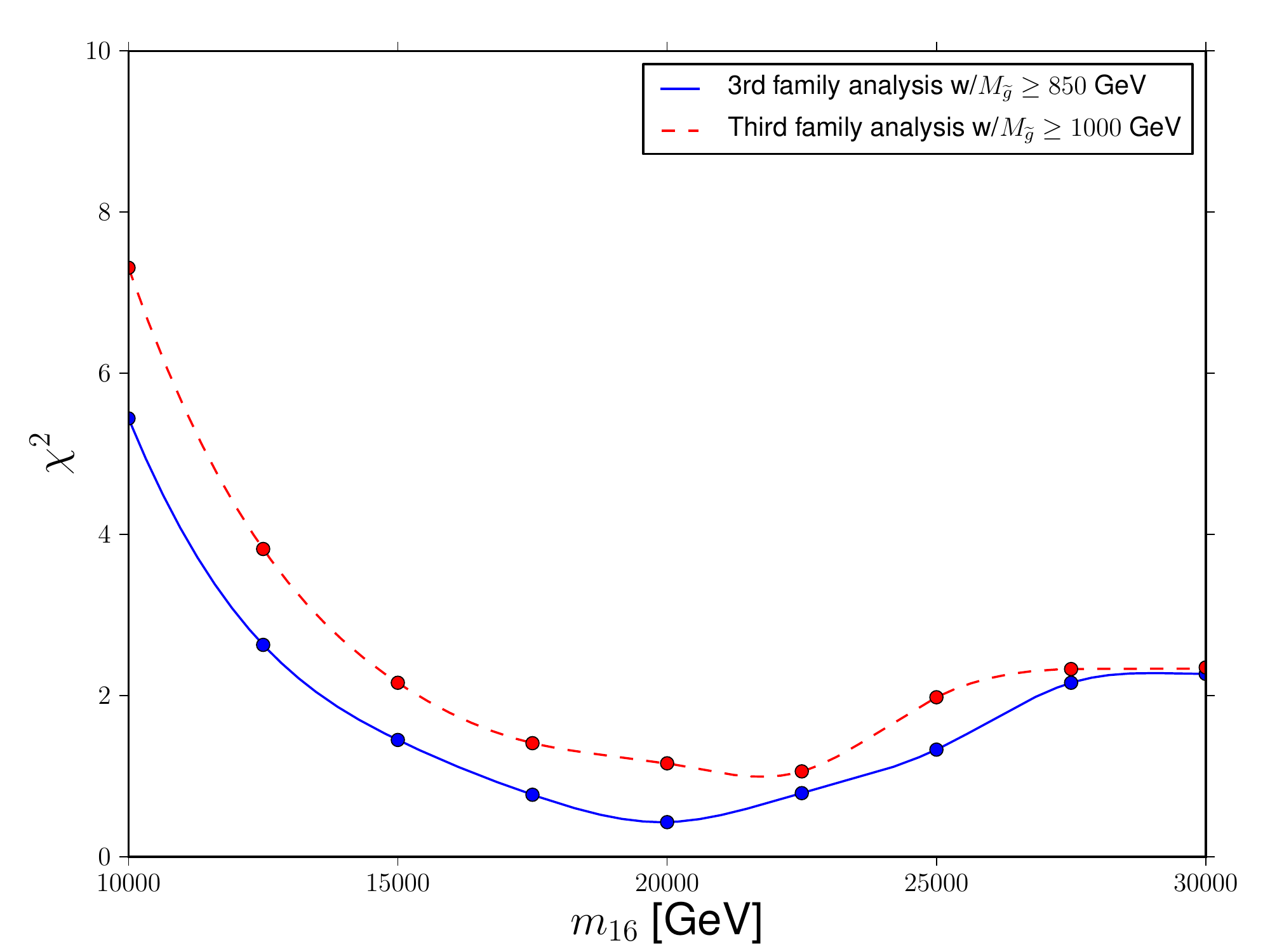}
\label{fig:thirdfamilychi2}
}\hspace{4ex}
\subfigure[\footnotesize As we increase the lower bound on the gluino mass, we find that $\chi^2$ dramatically increases for constant $m_{16}$.]{
\includegraphics[width=0.45\textwidth]{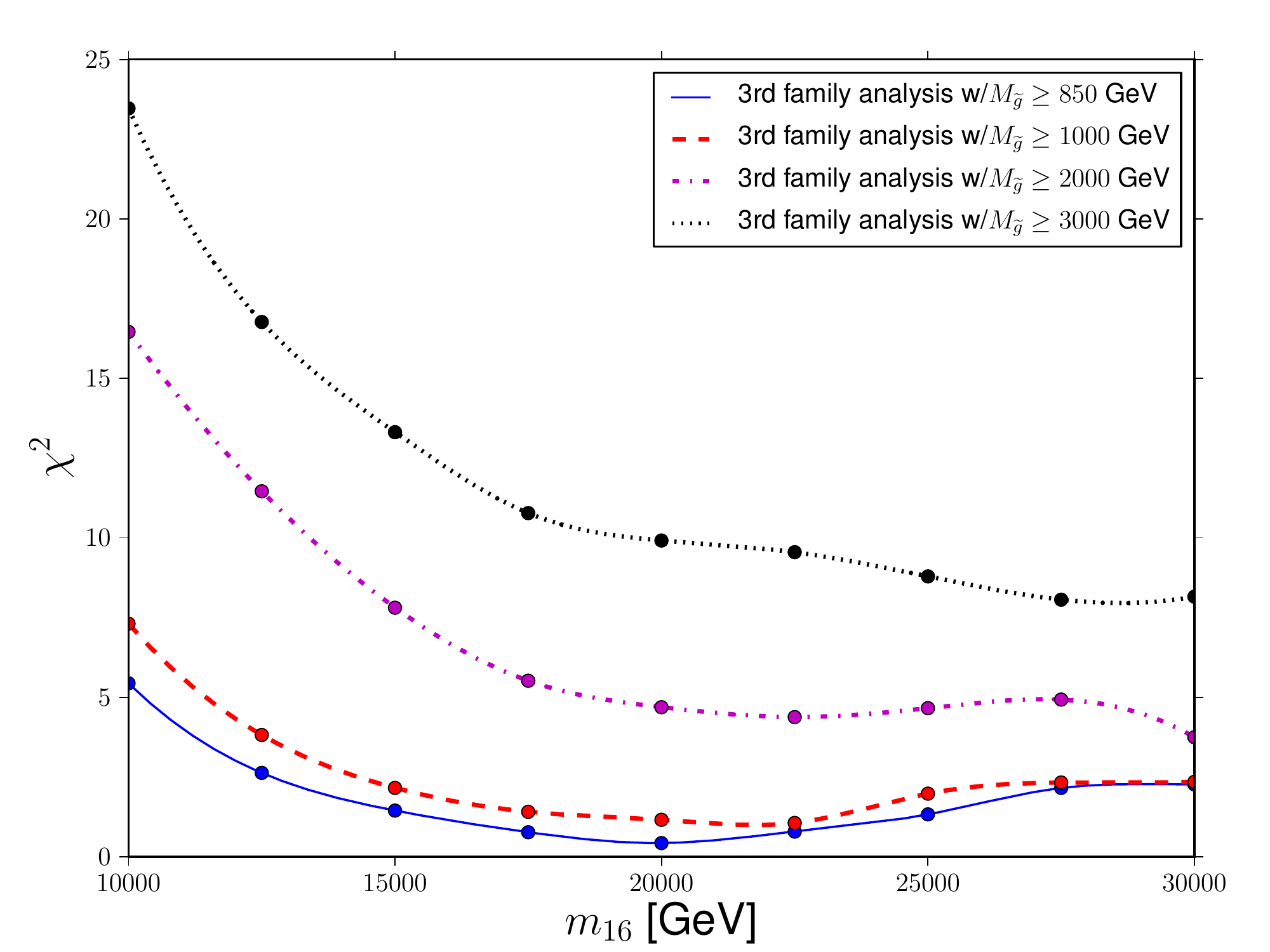}
\label{fig:gluinobounds}
}
\caption{\footnotesize $\chi^2$ vs.~$m_{16}$ for the third family analysis. The red and blue filled circles correspond to minima of $\chi^2$ for the values of $m_{16}$ indicated on the $x$-axis, where we have interpolated between them using cubic splines for ease of inspection. The curves correspond to different lower bounds on the gluino mass.}
\label{fig:thirdfamilychi2both}
\end{figure}

\begin{figure}[h!]
\centering
\includegraphics[width=0.7\textwidth]{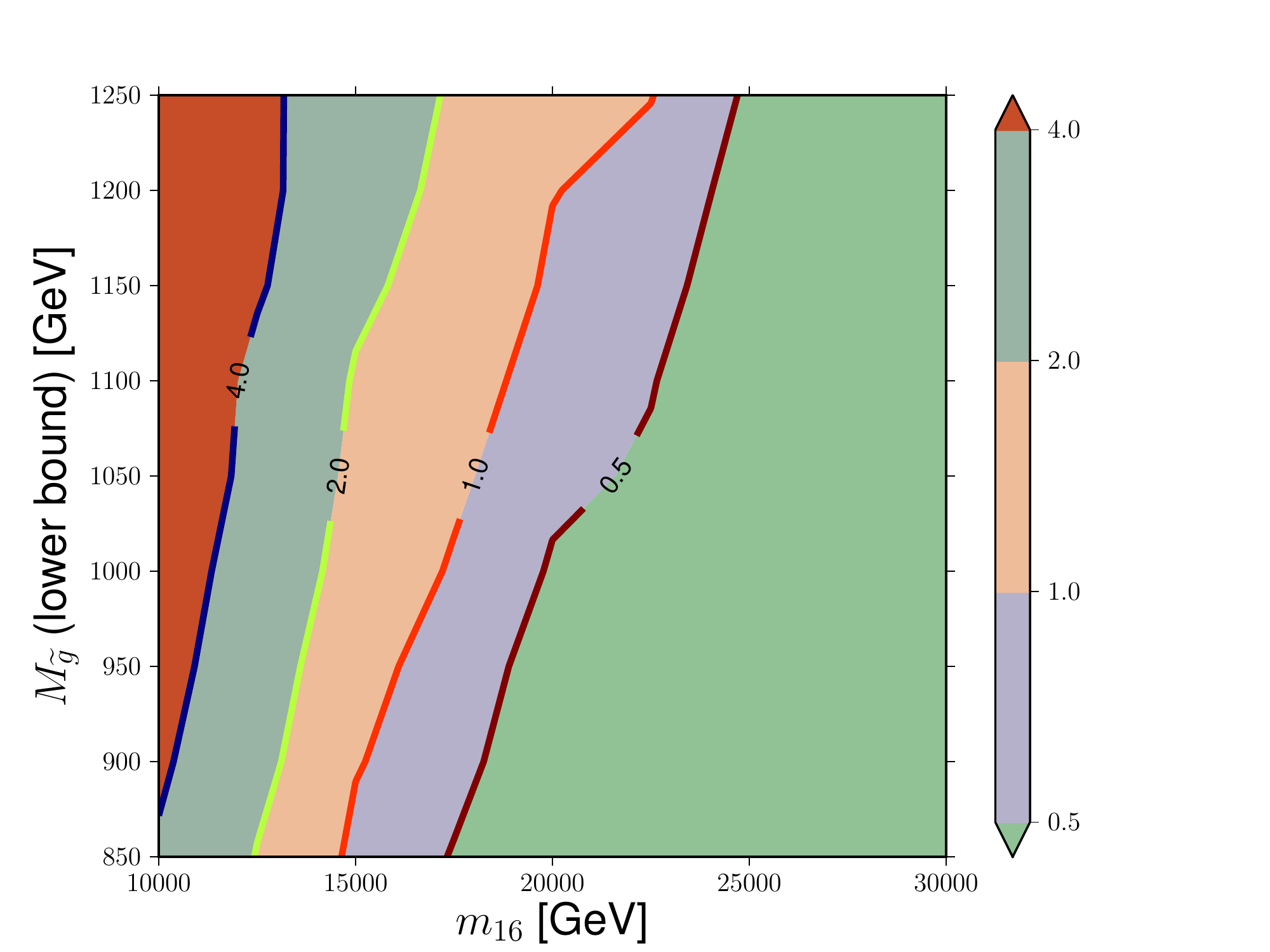}
\caption{\footnotesize The contribution of $M_h$ to the global $\chi^2$ as a function of the lower bound on the gluino mass (vertical axis) and the value of $m_{16}$ (horizontal axis). The Higgs mass is mainly responsible for the steep increase of $\chi^2$ observed in \ref{fig:thirdfamilychi2}.
}
\label{fig:mh_pull}
\end{figure}
\begin{figure}[h!]
\subfigure{
\includegraphics[width=0.55\textwidth]{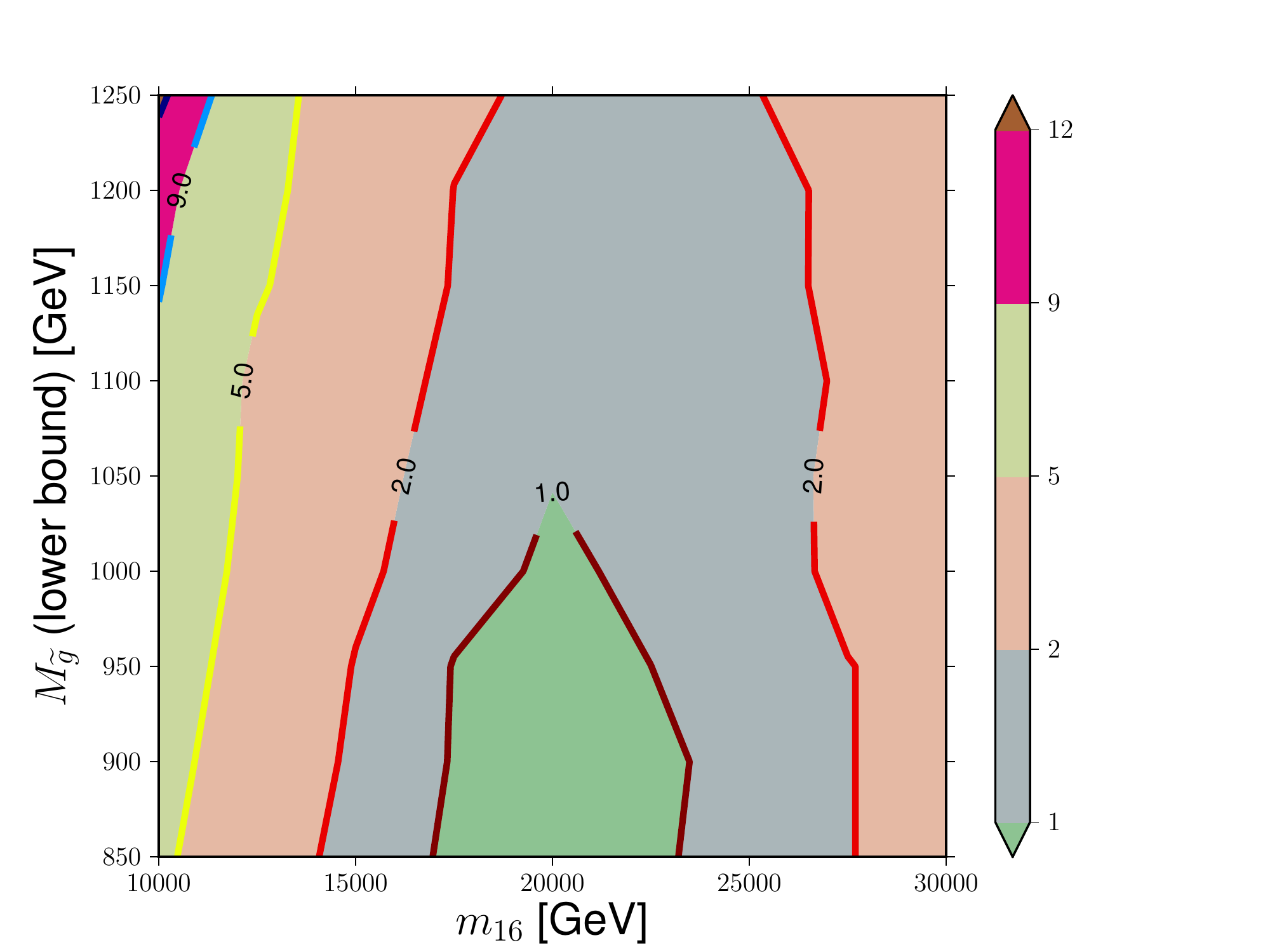}
\label{fig:chi2contours}
}
\subfigure{
\includegraphics[width=0.5\textwidth]{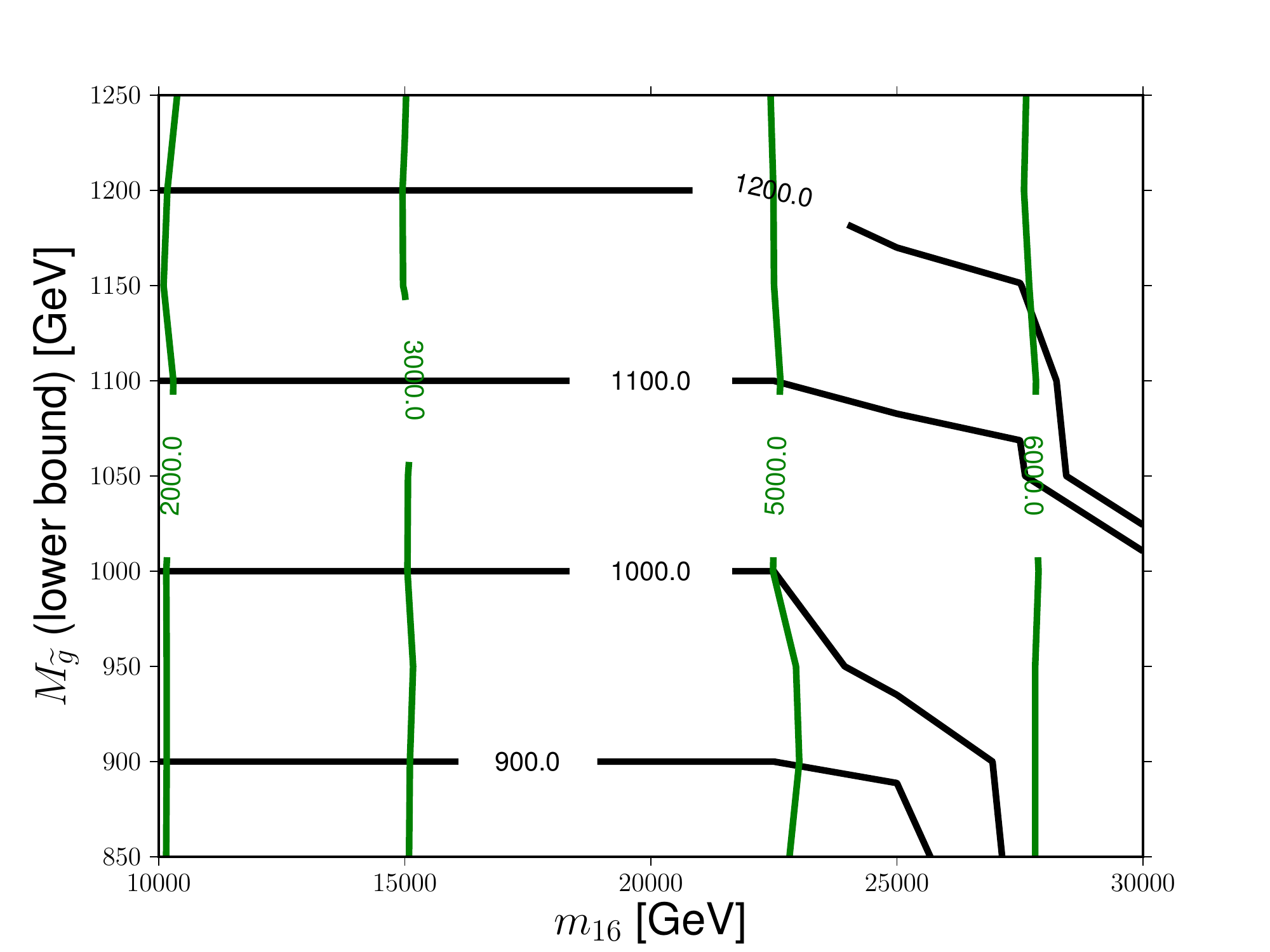}
\label{fig:mass_contours}
}
\caption{\footnotesize The best $\chi^2$ fits for the third family analysis (left) and contours of constant gluino mass ( roughly horizontal lines) and stop masses (vertical lines) (right).  Note, for larger values of $m_{16} \geq 22$ TeV, the best fit gluino pole mass is always much larger than the lower bound imposed.
}
\end{figure}

Consider first the SUSY spectrum in our analysis.  The first and
second family squarks and sleptons have mass of order $m_{16}$, while
stops, sbottoms and staus are all significantly lighter.  This is the
inverted scalar mass hierarchy which is a direct result of RG running.
Nevertheless, gluinos are always lighter than the third family squarks
and sleptons, and the lightest charginos and neutralinos are even
lighter.  Note the states $\tilde \chi^{\pm}$ and $\chi^0_2$ are
approximately degenerate.  A detailed spectrum is given in
\ref{spectrum}.  Recent results from CMS and ATLAS give lower bounds
on the gluino mass.  These bounds are given in terms of the CMSSM or
simplified models.  The simplified models which are most relevant for
our analysis are those in which (a) the third family of squarks and
sleptons are lighter than the first two, and (b) the gluino is lighter
than the stops and sbottoms.  In this case, the lower bound on the
gluino mass is now of order 1 - 1.2 TeV, {\it assuming} the branching
ratio $BR(\tilde g \rightarrow t \bar t \ \tilde \chi^0_{1}) = 100\%$
or $BR(\tilde g \rightarrow b \bar b \ \tilde \chi^0_{1}) = 100\%$
\cite{CMS-PAS-SUS-12-028, CMS-PAS-SUS-12-029}.  Although neither
simplified model is appropriate for our model, we nevertheless impose
a lower bound on the gluino pole mass in order to be roughly
consistent with the latest LHC results.

In \ref{fig:thirdfamilychi2} we present the best $\chi^2$ fits as a function of $m_{16}$ for two values of the lower bound
that we impose on the gluino pole mass, i.e.  850 and 1000 GeV.    We note that $\chi^2$ is relatively insensitive to these lower bounds on the gluino mass, although lower values of $M_{\tilde g}$ are slightly favored.  The minimum $\chi^2$ is found for $m_{16} = 20$ TeV, and $\chi^2$ increases as $m_{16}$ either decreases or increases. Features of the model like the large A-terms and large $\tan\beta$ are favorable to obtain a Higgs mass in the range of 122 - 127 GeV as observed at the LHC. However, the largest contribution to $\chi^2$ for lower values of $m_{16}$ comes from the Higgs mass constraint (see \ref{fig:mh_pull}).

As the lower bound on the gluino mass is increased to 2 or 3 TeV, we find that $\chi^2$ dramatically increases (see \ref{fig:gluinobounds}).   Note, this is predominantly due to the constraint from the light Higgs mass (\ref{fig:mh_pull}).  The simple explanation for this fact is that as the gluino mass increases the magnitude of $A_t$ at $M_{\mathrm SUSY}$ also increases, due to the infra-red fixed point.  This has the effect of decreasing the light Higgs mass because now $X_t > \sqrt{6} M_{\mathrm{SUSY}}$ which goes beyond maximal mixing.  As a consequence,  there appears to be an upper bound on the gluino mass of order 2 TeV, which makes gluinos inevitably observable at the LHC 14 TeV. However, as discussed earlier, the usual simplified models do not apply since gluinos decay with branching ratios  $\tilde g \rightarrow t \bar t \ \tilde \chi^0_{(1,2)}$, $b \bar b \ \tilde \chi^0_{(1,2)}$, $t \bar b \ \tilde \chi^-_{(1,2)}$, $b \bar t \ \tilde \chi^+_{(1,2)}$, $g \ \tilde \chi^0_{(1,2,3,4)}$ which are all significant.

In \ref{fig:chi2contours} we give the best $\chi^2$ fits for the third family analysis as a function of the lower bound on the gluino mass
and the value of $m_{16}$.  In \ref{fig:mass_contours} we give the contours of constant gluino masses (roughly horizontal lines) and stop masses (vertical lines)\footnote{In a recent analysis \cite{Elor:2012ig}, the authors found an upper bound on the stop mass for good Yukawa unification. Their result is a consequence of the constraint $\mu < 1000$ GeV, in order to satisfy dark matter bounds. We do not make any such assumption and do not find an upper bound on the stop mass for $m_{16} \le 30 $ TeV.}.  Note, for values of $m_{16} \geq 20$ TeV, the best fit gluino pole mass is always much larger than the lower bound imposed.

\section{Full Three family analysis \label{three family}}

\begin{figure}[h!]
\centering
\includegraphics[width=0.8\textwidth]{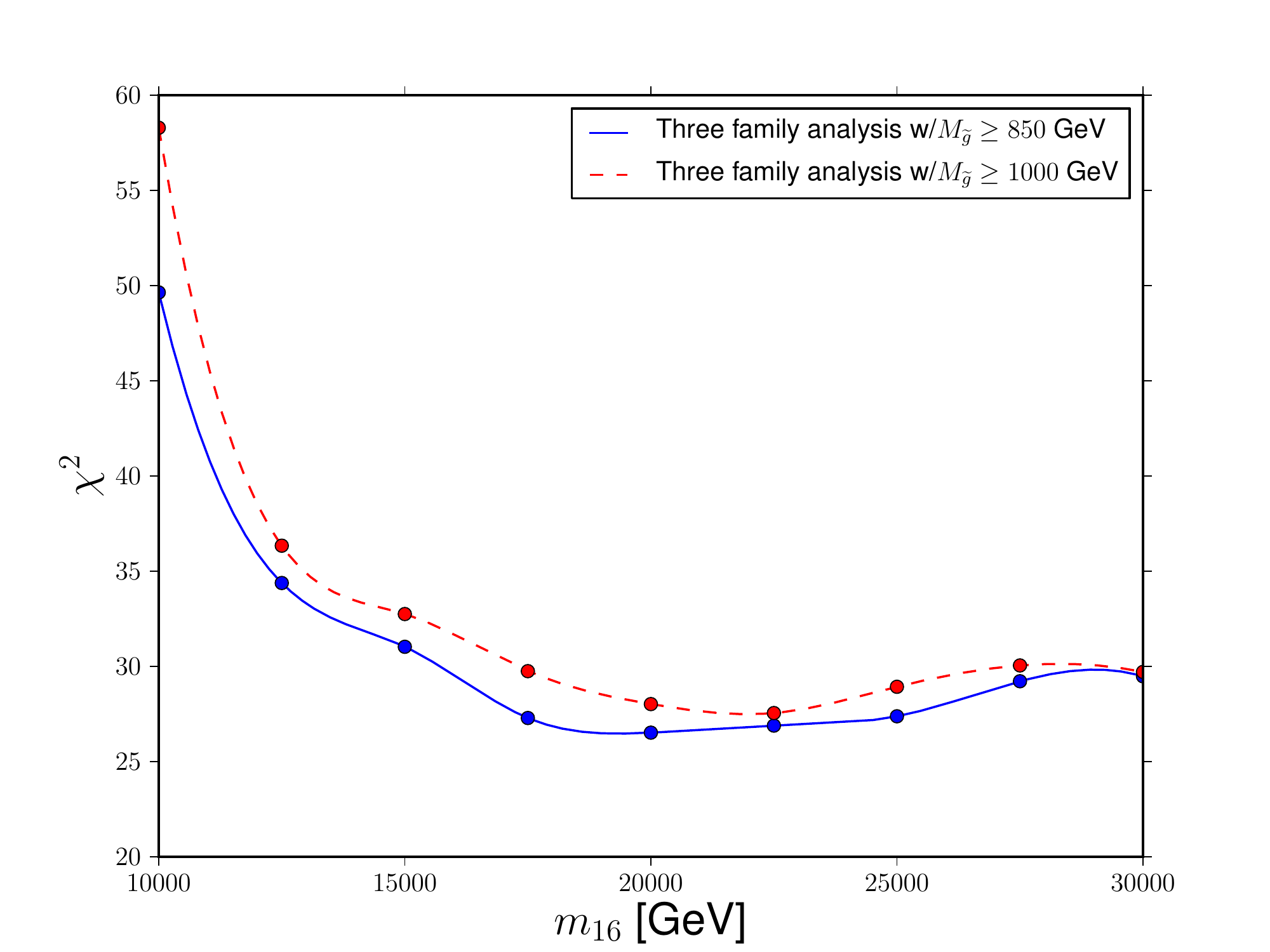}
\caption{\footnotesize $\chi^2$ vs.~$m_{16}$ for the full three family analysis. $\chi^2$ is very large for low values of $m_{16}\lesssim 15$ TeV. The blue (solid) and red (dashed) lines correspond to a lower bound on the gluino mass of 850 GeV and 1000 GeV, respectively.
}
\label{fig:3family}
\end{figure}

In this section we present the global $\chi^2$ analysis for three families including all 24 arbitrary parameters.  The $\chi^2$ function
includes 36 observables.   We present our results for fixed values of $m_{16}$ in \vref{fig:3family} and in \ref{t:fit10tev} to \ref{t:fit30tev}.

In \ref{fig:3family} we give the best $\chi^2$ fits for two different values of the lower bound on the gluino pole mass imposed
in the analysis. \ref{fig:thirdfamilychi2both} and \ref{fig:3family} have similar behavior.  The value of $\chi^2$ increases dramatically for values of $m_{16} \lesssim 15$ TeV.  For larger values of $m_{16} \gtrsim 25$ TeV the increase is much slower.
In the three family analysis, the minimum $\chi^2$ occurs around $m_{16} \approx 20$ TeV, just as in the third family analysis.  Moreover, the input
parameters which minimize $\chi^2$ in the third family analysis also minimize $\chi^2$ for the full three family analysis.

In \ref{t:fit10tev} to \ref{t:fit30tev} we present the best $\chi^2$ fits for values of $m_{16} = 10, \ 15, \ 20,  \  25, \ {\rm and} \  30$ TeV, respectively.  The best fit overall comes for $m_{16} = 20$ TeV with $\chi^2$/d.o.f $\sim 2$ (see \ref{t:fit20tev}).  Let us just comment on
a few of the initial values of the parameters for this point.   We find $\alpha_G \approx 1/26, \ M_G \approx 3 \times 10^{16} \ {\rm GeV}, \ {\rm and} \ \epsilon_3 = -1.45 \%$.   The magnitude of the Yukawa couplings are hierarchical. As expected, we have $A_0 \approx - 2 m_{16}$, $\mu, \ M_{1/2} \ll m_{16}$ and $\tan\beta \sim 50$.  The average Higgs mass parameter and the relative splitting are given by $m_{10} \equiv \sqrt{(m_{H_u}^2 + m_{H_d}^2)/2} \approx 26$ TeV and $\Delta_{m_H}^2 \equiv (m_{H_d}^2 - m_{H_u}^2)/(2 m_{10}^2) \approx 0.07$, respectively.  The gravitino mass for this model is expected to be of order the largest scalar mass, i.e. $M_{3/2} \sim m_{10} \approx 26$ TeV. We used \texttt{micrOmegas}\cite{Belanger:2010pz} to calculate the relic abundance for the benchmark points considered and found $\Omega^{\mathrm{th}} = 22.2$ (10 TeV), $\Omega^{\mathrm{th}} = 0.776$ (15 TeV), $\Omega^{\mathrm{th}} = 70.0$ (20 TeV), $\Omega^{\mathrm{th}} = 90.2$ (25 TeV), $\Omega^{\mathrm{th}} = 123$ (30 TeV). This is a consequence of a purely bino-like LSP.  In this case, a non-thermal process would be necessary to accommodate the observed dark matter abundance.  Assuming the correct dark matter abundance, a bino LSP would not have been observed yet by direct detection methods, but should be observable by future detectors \cite{Cheung:2012qy}.

Let us now focus on the fit.
Consider the observables with the largest pulls.  Roughly half the contribution to $\chi^2$ at this point comes from just two observables, namely $m_d/m_s$ and $\sin2\beta$.  Our value of $m_d/m_s$ is larger than the experimental value, and this implies that our value of $m_u/m_d \sim 0.9$ (see \ref{ellipse}).
We have allowed $|V_{ub}|$ to range over values consistent with both exclusive and inclusive measurements.   We find that our fit is more consistent with exclusive measurements.   Moreover, our fit value of $\sin2\beta$ is at the $3\sigma$ lower bound allowed
by the experiments.    Otherwise we are able to fit an amazing array of experimental observables.  The light Higgs mass is fit to within the $\pm$ 3 GeV theoretical uncertainty we have assigned.   As for the neutrino mixing angle $\theta_{13}$ we obtain a value closer to 6$^\circ$, rather than the present experimental value of approximately 9$^\circ$.   This may be a problem,
however, it has been noticed recently that flavor violating corrections to the K\"{a}hler potential can have a significant effect on $\theta_{13}$ without affecting the other larger mixing angles \cite{Chen:2012ha}.  Our neutrino spectrum corresponds to the normal hierarchy.  Note that the two large mixing angles are a consequence of a hierarchy in the right-handed neutrino masses.

\begin{table}
\centering
\begin{tabular}{|l|l|l|l|l|l|}
\hline
$m_{16}  $  &  10 TeV  &  15 TeV &  20 TeV & 25 TeV & 30 TeV \\
\hline
$\chi^2$ & 49.65 & 31.02 & 26.58 & 27.93 & 29.48  \\
\hline
$M_A$ & 2333 & 3662 & 1651 & 2029 &  2036  \\
$m_{\tilde t_1}$ & 1681 & 2529 & 3975 & 4892 & 5914   \\
$m_{\tilde b_1}$  & 2046 & 2972 & 5194 & 6353 & 7660   \\
$m_{\tilde \tau_1}$  & 3851 & 5576 & 7994 & 9769 &  11620   \\
$m_{\tilde\chi^0_1}$   & 133 & 134 & 137 & 149 &  167     \\
$m_{\tilde\chi^+_1}$  & 260 & 263 & 279 & 309 &  351     \\
$M_{\tilde g}$   & 853 & 850 & 851 & 910 &  1004    \\
\hline
\end{tabular}
\caption{\footnotesize SUSY Spectrum corresponding to the benchmark points presented in \ref{longtables}. The first two generation scalars have mass of the order of $m_{16}$.} \label{spectrum}
\end{table}

\begin{table}
 \centering
\begin{tabular}{|l|l|l|l|l|l|l|}
\hline
& Current Limit & 10 TeV & 15 Te V & 20 TeV & 25 TeV & 30 TeV \\
\hline
e EDM $\times 10^{28}$ & $< 10.5 \ e\ cm$ & $-0.224$ & $-0.0408 $ & $-0.0173 $ & $-0.0113 $ & $-0.0084 $\\
$\mu$ EDM $\times 10^{28}$& $(-0.1 \pm 0.9) \times 10^{9}\ e\ cm$ & $34.6 $ & $6.23 $& $3.04 $ & $1.77 $ & $1.20$\\
$\tau$ EDM $\times 10^{28}$ & $-0.220 - 0.45 \times 10^{12}\ e\ cm$ & $-2.09 $ & $-0.394 $& $-0.185 $ & $-0.109 $ & $-0.0732 $\\
\hline
BR$(\mu \rightarrow e \gamma) \times 10^{12}$ & $< 2.4 \ e\ cm$ & $5.09 $ &$1.23 $ & $0.211 $ & $0.0937 $ & $0.0447 $\\
BR$(\tau \rightarrow e \gamma) \times 10^{12}$  & $< 3.3 \times 10^{4}\ e\ cm$ & $58.8 $ &$13.9 $ & $2.40 $ & $1.04 $ &$0.502 $ \\
BR$(\tau \rightarrow \mu \gamma) \times 10^{8}$  & $< 4.4 \ e\ cm$ &$1.75 $ &$0.498 $ & $0.0837 $& $0.0385 $& $0.0182 $\\
\hline
sin $\delta$ & & -0.60 & -0.87 & -0.27 & -0.42 & -0.53 \\
\hline
\end{tabular}
\caption{\footnotesize Predictions from the full three family analysis. The dipole moments and branching ratios were calculated using \susyflavor.} \label{predictions}
\end{table}

In \ref{spectrum} we summarize the predictions for the SUSY spectrum given values of $m_{16} = 10, 15, 20, 25, 30$ TeV, respectively.  We give the spectrum of the lightest squark, slepton and gaugino masses, and the CP odd Higgs mass $M_A$.  The first and second generation squarks and sleptons all have mass of order $m_{16}$.   Note, in order
to fit the branching ratio $BR(B_s \rightarrow \mu^+ \ \mu^-)$ with large $\tan\beta$, we have $M_A \gg M_Z$.  Thus we are in the decoupling limit where the light Higgs is predicted to couple to matter just like the Standard Model Higgs.   Therefore, any deviation from this prediction would rule out our model.     Finally, in \ref{predictions} we present results for yet to be observed quantities such as electric dipole moments of charged leptons, flavor violating processes such as $BR(\mu \rightarrow e \ \gamma)$ and the CP violating angle in the lepton sector, $\sin\delta$.   The value of $\sin\delta$ is close to zero and is thus consistent with tentative emerging hints for $\delta\simeq\pi$ \cite{Fogli:2012ua}.  We also find that the $BR(\mu \rightarrow e \ \gamma)$ may in fact be observable by the MEG experiment in a few years \cite{Mihara:2012zz}.

\section{Summary and Conclusions \label{conclusion}}

We have performed a global $\chi^2$ analysis of an \SO{10} SUSY GUT times a $D_3 \times [\U{1} \times \mathbb{Z}_2 \times \mathbb{Z}_3]$ family symmetry.  The model fits all fermion masses and mixing angles, as well as many flavor observables, quite well.  The model has 24 arbitrary parameters which we use to fit
36 low energy observables. Five of these parameters include the soft SUSY breaking masses,  a universal squark and slepton mass, $m_{16}$; a universal
cubic scalar coupling, $A_0$;  a universal gaugino mass, $M_{1/2}$ and split Higgs up and down masses, $m_{H_u}, \ m_{H_d}$.   The model has gauge coupling
unification and top, bottom, $\tau$, $\nu_\tau$ Yukawa unification at $M_{GUT}$.
We have analyzed the model for the third family alone and then for three families.  We have shown that the SUSY spectrum is predominantly determined by fitting the third family and light Higgs masses and the branching ratio $BR(B_s \rightarrow \mu^+ \ \mu^-)$.  In \ref{t:fit10tev} - \ref{t:fit30tev} we give the best three family fits for five different values of the universal scalar mass $m_{16}$.  The best overall fit is found for $m_{16} \approx 20$ TeV.  The SUSY spectrum for these best fit points are given in \ref{spectrum}.

Our model makes several significant predictions.
\begin{enumerate}[(i)]
\item The first and second family of squarks and sleptons obtain mass of order $m_{16}$, while the third family scalars are naturally much lighter.   Then gluinos and the lightest chargino and neutralinos are always lighter than the third family squarks and sleptons.
\item Due to Yukawa unification of the third family at the GUT scale we have $\tan\beta \approx 50$.   In order to fit the branching ratio $BR(B_s \rightarrow \mu^+ \ \mu^-)$ we find the CP odd Higgs mass, $m_A \gg M_Z$.   Hence we are in the decoupling limit and the light Higgs is predicted to be Standard Model-like.
\item In order to fit the light Higgs mass, we find an upper bound on the gluino mass, $M_{\tilde g} \sim 2$ TeV.  Thus gluinos should be observable
at LHC14.
\item  No simplified model studied to date describes the relevant gluino decay branching ratios (See scenarios studied in \cite{Toharia:2005gm}).   Thus in order to constrain our theory we need both
CMS and ATLAS to provide detailed bounds on the $p \ p \rightarrow \tilde g \ \tilde g$  production cross-section times branching ratios for the many
different two and three body decay modes, i.e. $\tilde g \rightarrow t \bar t \ \tilde \chi^0_{(1,2)};$ $b \bar b \ \tilde \chi^0_{(1,2)};$ $t \bar b \ \tilde \chi^-_{(1,2)};$ $b \bar t \ \tilde \chi^+_{(1,2)};$ $g \ \tilde \chi^0_{(1,2,3,4)}$.\footnote{In order for ATLAS or CMS to test our model, we would gladly provide an SLHA2 file specifying all the low energy parameters of any of our benchmark points.}
\item  We find $BR(\mu \rightarrow e \ \gamma) \sim 10^{-12} - 10^{-13}$ for values of $m_{16} = 15 - 25$ TeV.   This may soon be observable at MEG \cite{Mihara:2012zz}.
\item We find the CP violating parameter in the lepton sector,  $\sin\delta \approx 0$, and the neutrinos obey a normal hierarchy.
\item Since the first two family sleptons have mass of order $m_{16}$ we are not able to fit the muon anomalous magnetic moment, $(g-2)_\mu$.
\item Our LSP is predominantly bino and thus assuming a thermal calculation of the relic abundance, we find $\Omega_{\tilde \chi^0_1}$ too large.
\item The gravitino mass is naturally of order $\sqrt{2} \ m_{16}$ or for $m_{16} = 15 - 25$ TeV we have $m_{\tilde G} \sim 20 - 35$ TeV.  Thus the the model may avoid the cosmological gravitino problem.
\end{enumerate}

There is one obvious issue with the model regarding fine-tuning. We have not performed a detailed analysis of fine-tuning, but
a rough measure is given by $\Delta = (\frac{\mu}{M_Z})^2 \sim 150$, corresponding to a fine-tuning of $1/\Delta$. As this is true for most of the surviving parameter space of the MSSM, at the moment
we do not regard this as a serious problem. The question of electroweak fine-tuning in Yukawa unified models was recently studied in \cite{Baer:2012jp}.

Let us now consider future directions. We will evaluate the gluino decay branching ratios in our model in order to compare to
LHC data in a future work. In addition, we want to analyze other boundary conditions at the GUT scale consistent with gauge and
Yukawa coupling unification. In particular, we will consider the ``DR3" scheme \cite{Baer:2009ie} and also non-universal gaugino masses as discussed in \cite{Badziak:2011wm}, and study them with the combined predictive power of family symmetries. On the computational front, we would like to explore other methods to tackle the problem of finding a global minimum in a multi-dimensional parameter space.


\section*{Acknowledgments}

We are indebted to Radovan Derm\'i\v{s}ek for his program and his valuable inputs in using it. We also thank Christopher Plumberg and Aditi Raval for their work in the early stages of the project. We acknowledge useful discussions with Athanasios Dedes, Chris Hill, Nazila Mahmoudi, and Janusz Rosiek. We thank Pietro Slavich for sharing their Higgs boson mass calculation routine. A.A.~and S.R.~received partial support for this work from DOE/ER/01545-896. A.W.~acknowledges partial support from LabEx ENIGMASS and would like to thank the Ohio State University for their hospitality. We thank the \emph{Ohio Supercomputer Center} and
the \emph{Centre de Calcul de l'Institut National de Physique
Nucl\'{e}aire et Physique des Particules} in Lyon for using their
resources.

\clearpage

\newpage

\appendix

\labelformat{section}{Appendix #1}

\section{Benchmark Points}
\label{longtables}

\begin{table}[h!]
\centering
\begin{footnotesize}
 \caption[8]{ \mbox{Initial parameters for benchmark point with $ {\boldsymbol{m_{16} = 10 \; {\rm TeV}}}$: }\\ (1/$\alpha_G, \, M_G, \, \epsilon_3$) = ($25.42,\, 2.80 \times 10^{16}$ GeV, $\, -2.20$ \%),\\
($\lambda, \, \lambda \epsilon, \, \sigma, \, \lambda \tilde \epsilon, \, \rho, \,
\lambda \epsilon', \, \lambda \epsilon \xi$) = ($ 0.61, \, 0.031, \, 1.14, \, 0.0048, \,
0.071, \, -0.0019, \, 0.0038  $),\\
($\Phi_\sigma, \, \Phi_{\tilde \epsilon}, \, \Phi_\rho, \,
\Phi_\xi$) =  ($0.517, \, 0.625, \, 4.000, \, 3.497$) rad,
\makebox[6.6em]{ }\\
($m_{16}, \, M_{1/2}, \, A_0, \, \mu(M_Z)$) = ($10000,\, 239, \,
-20247, \,
791.13$) GeV,\\
($(m_{H_d}/m_{16})^2, \, (m_{H_u}/m_{16})^2, \, \tan\beta$) = ($1.95,  \, 1.61, \, 49.42$) \\
($M_{R_3}, \, M_{R_2}, \, M_{R_1}$) = ($3.2 \times 10^{13}$ GeV, $\,
5.6 \times 10^{11} $ GeV, $\, 0.9 \times 10^{10} $ GeV)  \label{t:fit10tev} 
}
\end{footnotesize}

\vspace{5mm}
\scalebox{0.9}{
\begin{tabular}{|l|c|c|c|c|}
\hline
Observable  &  Fit value  &  Exp value  &  Pull & $\sigma$  \\
\hline
$M_Z$ &              91.1876         &  91.1876         &  0.0000          &  0.4559          \\
$M_W$ &              80.5581         &  80.3850         &  0.4305          &  0.4022          \\
$1/\alpha_{em}$ &    136.3909        &  137.0360        &  0.9415          &  0.6852          \\
$G_{\mu} \times 10^5$ & 1.1754          &  1.1664          &  0.7722          &  0.0117          \\
$\alpha_3$ &         0.1184          &  0.1184          &  0.0342          &  0.0009          \\
\hline
$M_t$ &              173.9306        &  173.5000        &  0.3253          &  1.3238          \\
$m_b(m_b)$ &         4.1719          &  4.1800          &  0.2213          &  0.0366          \\
$M_{\tau}$ &         1.7796          &  1.7768          &  0.3104          &  0.0089          \\
\hline
$m_c(m_c)$ &         1.2782          &  1.2750          &  0.1246          &  0.0258          \\
$m_s$ &              0.0962          &  0.0950          &  0.2437          &  0.0050          \\
$m_d/m_s$  &         0.0710          &  0.0526          &  3.3115          &  0.0055          \\
$1/Q^2$ &            0.0019          &  0.0019          &  0.2854          &  0.0001          \\
$M_{\mu}$ &          0.1056          &  0.1057          &  0.2011          &  0.0005          \\
$M_e \times 10^4$ &  5.1137          &  5.1100          &  0.1450          &  0.0255          \\
\hline
$|V_{us}|$ &         0.2248          &  0.2252          &  0.2886          &  0.0014          \\
$|V_{cb}|$ &         0.0424          &  0.0406          &  0.8748          &  0.0020          \\
$|V_{ub}| \times 10^3$ & 3.3462          &  3.7700          &  0.4985          &  0.8502          \\
$|V_{td}| \times 10^3$ & 9.5185          &  8.4000          &  1.8597          &  0.6015          \\
$|V_{ts}|$ &         0.0414          &  0.0429          &  0.5683          &  0.0026          \\
$\text{sin} 2\beta$ & 0.6357          &  0.6790          &  2.1346          &  0.0203          \\
\hline
$\epsilon_K$ &       0.0023          &  0.0022          &  0.2568          &  0.0002          \\
$\Delta M_{Bs}/\Delta M_{Bd}$ & 59.6805         &  35.0600         &  3.5049          &  7.0246          \\
$\Delta M_{Bd} \times 10^{13}$ & 3.5432          &  3.3370          &  0.3086          &  0.6682          \\
\hline
$m^2_{21} \times 10^5$ &   7.6408          &  7.5450          &  0.1754          &  0.5463          \\
$m^2_{31} \times 10^3$ &  2.6521          &  2.4800          &  0.8182          &  0.2104          \\
$\text{sin}^2 \theta_{12}$ & 0.3297          &  0.3050          &  0.7041          &  0.0350          \\
$\text{sin}^2 \theta_{23}$ & 0.6441          &  0.5050          &  0.8427          &  0.1650          \\
$\text{sin}^2 \theta_{13}$ & 0.0128          &  0.0230          &  1.4585          &  0.0070          \\
\hline
$M_h$ &              116.94          &  125.30          &  2.7265          &  3.0676          \\
\hline
$BR(B \rightarrow X_s \gamma) \times 10^4$ & 3.9408          &  3.4300          &  0.3120          &  1.6374          \\
$BR(B_s \rightarrow \mu^+ \mu^-) \times 10^9$ & 2.3710          &  3.2000          &  0.5083          &  1.6308          \\
$BR(B_d \rightarrow \mu^+ \mu^-) \times 10^{10}$ & 1.7509          &  8.1000          &  0.0000          &  5.2559          \\
$BR(B \rightarrow \tau \nu) \times 10^5$ & 7.1988          &  16.6000         &  1.0525          &  8.9320          \\
$BR(B \rightarrow K^*\mu^+ \mu^-)$(low) $\times 10^8$& 5.4370          &  19.7000         &  1.1881          &  12.0051         \\
$BR(B \rightarrow K^*\mu^+ \mu^-)$(high) $\times 10^8$ & 7.8844          &  12.0000         &  0.5651          &  7.2835          \\
$q_0^2(B \rightarrow K^* \mu^+ \mu^-)$ & 4.8731          &  4.9000          &  0.0206          &  1.3009          \\
\hline
\multicolumn{3}{|l}{Total $\chi^2$}  &  \textbf{49.6463}&  \\\hline
\end{tabular}}
\end{table}

\begin{table}
\centering
\begin{footnotesize}
\caption[8]{ \mbox{Initial parameters for benchmark point with $ {\boldsymbol{m_{16} = 15 \; {\rm TeV}}}$: }\\ (1/$\alpha_G, \, M_G, \, \epsilon_3$) = ($25.50,
\, 2.96 \times 10^{16}$ GeV, $\, -2.40$ \%), \makebox[1.8em]{ }\\
($\lambda, \, \lambda \epsilon, \, \sigma, \, \lambda \tilde \epsilon, \, \rho, \,
\lambda \epsilon', \, \lambda \epsilon \xi$) = ($ 0.61, \, 0.031, \, 1.14, \, 0.0049, \,
0.070, \, -0.0019, \, 0.0037  $),\\
($\Phi_\sigma, \, \Phi_{\tilde \epsilon}, \, \Phi_\rho, \,
\Phi_\xi$) =  ($0.527, \, 0.635, \, 3.881, \, 3.429$) rad,
\makebox[6.6em]{ }\\
($m_{16}, \, M_{1/2}, \, A_0, \, \mu(M_Z)$) = ($15000,\, 201, \,
-30639, \,
513.07$) GeV,\\
($(m_{H_d}/m_{16})^2, \, (m_{H_u}/m_{16})^2, \, \tan\beta$) = ($1.97,  \, 1.62, \, 49.59$) \\
($M_{R_3}, \, M_{R_2}, \, M_{R_1}$) = ($4.2 \times 10^{13}$ GeV, $\,
6.1 \times 10^{11} $ GeV, $\, 1.0 \times 10^{10} $ GeV) \label{t:fit15tev} 
}
\end{footnotesize}
 \vspace{5mm}
\begin{tabular}{|l|c|c|c|c|}
\hline
Observable  &  Fit value  &  Exp value  &  Pull & Sigma  \\
\hline
$M_Z$ &              91.1876         &  91.1876         &  0.0000          &  0.4559          \\
$M_W$ &              80.5671         &  80.3850         &  0.4527          &  0.4022          \\
$1/\alpha_{em}$ &    136.4172        &  137.0360        &  0.9031          &  0.6852          \\
$G_{\mu} \times 10^5$ & 1.1766          &  1.1664          &  0.8739          &  0.0117          \\
$\alpha_3$ &         0.1185          &  0.1184          &  0.1342          &  0.0009          \\
\hline
$M_t$ &              173.5253        &  173.5000        &  0.0191          &  1.3238          \\
$m_b(m_b)$ &         4.1903          &  4.1800          &  0.2813          &  0.0366          \\
$M_{\tau}$ &         1.7756          &  1.7768          &  0.1366          &  0.0089          \\
\hline
$m_c(m_c)$ &         1.2613          &  1.2750          &  0.5312          &  0.0258          \\
$m_s$ &              0.0964          &  0.0950          &  0.2766          &  0.0050          \\
$m_d/m_s$  &         0.0686          &  0.0526          &  2.8819          &  0.0055          \\
$1/Q^2$ &            0.0018          &  0.0019          &  0.4900          &  0.0001          \\
$M_{\mu}$ &          0.1056          &  0.1057          &  0.0748          &  0.0005          \\
$M_e \times 10^4$ &  5.1135          &  5.1100          &  0.1386          &  0.0255          \\
\hline
$|V_{us}|$ &         0.2243          &  0.2252          &  0.6542          &  0.0014          \\
$|V_{cb}|$ &         0.0410          &  0.0406          &  0.1681          &  0.0020          \\
$|V_{ub}| \times 10^3$ & 3.1115          &  3.7700          &  0.7745          &  0.8502          \\
$|V_{td}| \times 10^3$ & 8.8886          &  8.4000          &  0.8124          &  0.6015          \\
$|V_{ts}|$ &         0.0401          &  0.0429          &  1.0638          &  0.0026          \\
$\text{sin} 2\beta$ & 0.6220          &  0.6790          &  2.8094          &  0.0203          \\
\hline
$\epsilon_K$ &       0.0023          &  0.0022          &  0.1079          &  0.0002          \\
$\Delta M_{Bs}/\Delta M_{Bd}$ & 37.6694         &  35.0600         &  0.3715          &  7.0246          \\
$\Delta M_{Bd} \times 10^{13}$ & 4.0059          &  3.3370          &  1.0010          &  0.6682          \\
\hline
$m^2_{21} \times 10^5$ &   7.5155          &  7.5450          &  0.0540          &  0.5463          \\
$m^2_{31} \times 10^3$ &  2.5097          &  2.4800          &  0.1413          &  0.2104          \\
$\text{sin}^2 \theta_{12}$ & 0.2994          &  0.3050          &  0.1600          &  0.0350          \\
$\text{sin}^2 \theta_{23}$ & 0.7414          &  0.5050          &  1.4323          &  0.1650          \\
$\text{sin}^2 \theta_{13}$ & 0.0147          &  0.0230          &  1.1908          &  0.0070          \\
\hline
$M_h$ &              122.21          &  125.30          &  1.0080          &  3.0676          \\
\hline
$BR(B \rightarrow X_s \gamma) \times 10^4$ & 3.5456          &  3.4300          &  0.0706          &  1.6374          \\
$BR(B_s \rightarrow \mu^+ \mu^-) \times 10^9$ & 4.3688          &  3.2000          &  0.7167          &  1.6308          \\
$BR(B_d \rightarrow \mu^+ \mu^-) \times 10^{10}$ & 1.3486          &  8.1000          &  0.0000          &  5.2559          \\
$BR(B \rightarrow \tau \nu) \times 10^5$ & 6.2875          &  16.6000         &  1.1546          &  8.9320          \\
$BR(B \rightarrow K^*\mu^+ \mu^-)$(low) $\times 10^8$& 5.0499          &  19.7000         &  1.2203          &  12.0051         \\
$BR(B \rightarrow K^*\mu^+ \mu^-)$(high) $\times 10^8$ & 7.5449          &  12.0000         &  0.6117          &  7.2835          \\
$q_0^2(B \rightarrow K^* \mu^+ \mu^-)$ & 4.5922          &  4.9000          &  0.2366          &  1.3009          \\
\hline
\multicolumn{3}{|l}{Total $\chi^2$}  &  \textbf{31.0266}&  \\\hline
\end{tabular}
\end{table}

\begin{table}
\centering
\begin{footnotesize}
\caption[8]{ \mbox{Initial parameters for benchmark point with $ {\boldsymbol{m_{16} = 20 \; {\rm TeV}}}$: }\\ (1/$\alpha_G, \, M_G, \, \epsilon_3$) = ($25.90,
\, 3.13 \times 10^{16}$ GeV, $\, -1.45$ \%), \makebox[1.8em]{ }\\
($\lambda, \, \lambda \epsilon, \, \sigma, \, \lambda \tilde \epsilon, \, \rho, \,
\lambda \epsilon', \, \lambda \epsilon \xi$) = ($ 0.60, \, 0.031, \, 1.14, \, 0.0049, \,
0.070, \, -0.0019, \, 0.0038  $),\\
($\Phi_\sigma, \, \Phi_{\tilde \epsilon}, \, \Phi_\rho, \,
\Phi_\xi$) =  ($0.533, \, 0.548, \, 3.936, \, 3.508$) rad,
\makebox[6.6em]{ }\\
($m_{16}, \, M_{1/2}, \, A_0, \, \mu(M_Z)$) = ($20000,\, 168, \,
-41087, \,
1163.25$) GeV,\\
($(m_{H_d}/m_{16})^2, \, (m_{H_u}/m_{16})^2, \, \tan\beta$) = ($1.85,  \, 1.61, \, 49.82$) \\
($M_{R_3}, \, M_{R_2}, \, M_{R_1}$) = ($3.2 \times 10^{13}$ GeV, $\,
6.1 \times 10^{11} $ GeV, $\, 0.9 \times 10^{10} $ GeV)  \label{t:fit20tev} 
}
\end{footnotesize}
\vspace{5mm}
\begin{tabular}{|l|l|l|l|l|}
\hline
Observable  &  Fit value  &  Exp value  &  Pull & Sigma  \\
\hline
$M_Z$ &              91.1876         &  91.1876         &  0.0000          &  0.4559          \\
$M_W$ &              80.5452         &  80.3850         &  0.3982          &  0.4022          \\
$1/\alpha_{em}$ &    137.0725        &  137.0360        &  0.0533          &  0.6852          \\
$G_{\mu} \times 10^5$ & 1.1713          &  1.1664          &  0.4250          &  0.0117          \\
$\alpha_3$ &         0.1184          &  0.1184          &  0.0467          &  0.0009          \\
\hline
$M_t$ &              174.0184        &  173.5000        &  0.3916          &  1.3238          \\
$m_b(m_b)$ &         4.1849          &  4.1800          &  0.1334          &  0.0366          \\
$M_{\tau}$ &         1.7755          &  1.7768          &  0.1462          &  0.0089          \\
\hline
$m_c(m_c)$ &         1.2547          &  1.2750          &  0.7876          &  0.0258          \\
$m_s$ &              0.0964          &  0.0950          &  0.2807          &  0.0050          \\
$m_d/m_s$  &         0.0692          &  0.0526          &  2.9891          &  0.0055          \\
$1/Q^2$ &            0.0018          &  0.0019          &  0.4749          &  0.0001          \\
$M_{\mu}$ &          0.1056          &  0.1057          &  0.1049          &  0.0005          \\
$M_e \times 10^4$ &  5.1122          &  5.1100          &  0.0862          &  0.0255          \\
\hline
$|V_{us}|$ &         0.2243          &  0.2252          &  0.5964          &  0.0014          \\
$|V_{cb}|$ &         0.0415          &  0.0406          &  0.4511          &  0.0020          \\
$|V_{ub}| \times 10^3$ & 3.2023          &  3.7700          &  0.6678          &  0.8502          \\
$|V_{td}| \times 10^3$ & 8.9819          &  8.4000          &  0.9675          &  0.6015          \\
$|V_{ts}|$ &         0.0407          &  0.0429          &  0.8518          &  0.0026          \\
$\text{sin} 2\beta$ & 0.6304          &  0.6790          &  2.3959          &  0.0203          \\
\hline
$\epsilon_K$ &       0.0023          &  0.0022          &  0.3823          &  0.0002          \\
$\Delta M_{Bs}/\Delta M_{Bd}$ & 39.4933         &  35.0600         &  0.6311          &  7.0246          \\
$\Delta M_{Bd} \times 10^{13}$ & 3.9432          &  3.3370          &  0.9072          &  0.6682          \\
\hline
$m^2_{21} \times 10^5$ &   7.5126          &  7.5450          &  0.0593          &  0.5463          \\
$m^2_{31} \times 10^3$ &  2.4828          &  2.4800          &  0.0135          &  0.2104          \\
$\text{sin}^2 \theta_{12}$ & 0.2949          &  0.3050          &  0.2880          &  0.0350          \\
$\text{sin}^2 \theta_{23}$ & 0.5156          &  0.5050          &  0.0640          &  0.1650          \\
$\text{sin}^2 \theta_{13}$ & 0.0131          &  0.0230          &  1.4134          &  0.0070          \\
\hline
$M_h$ &              124.07          &  125.30          &  0.4010          &  3.0676          \\
\hline
$BR(B \rightarrow X_s \gamma) \times 10^4$ & 3.4444          &  3.4300          &  0.0088          &  1.6374          \\
$BR(B_s \rightarrow \mu^+ \mu^-) \times 10^9$ & 1.6210          &  3.2000          &  0.9682          &  1.6308          \\
$BR(B_d \rightarrow \mu^+ \mu^-) \times 10^{10}$ & 1.0231          &  8.1000          &  0.0000          &  5.2559          \\
$BR(B \rightarrow \tau \nu) \times 10^5$ & 6.3855          &  16.6000         &  1.1436          &  8.9320          \\
$BR(B \rightarrow K^*\mu^+ \mu^-)$(low) $\times 10^8$& 5.1468          &  19.7000         &  1.2123          &  12.0051         \\
$BR(B \rightarrow K^*\mu^+ \mu^-)$(high) $\times 10^8$ & 7.7469          &  12.0000         &  0.5839          &  7.2835          \\
$q_0^2(B \rightarrow K^* \mu^+ \mu^-)$ & 4.5168          &  4.9000          &  0.2945          &  1.3009          \\
\hline
\multicolumn{3}{|l}{Total $\chi^2$}  &  \textbf{26.5812}&  \\\hline
\end{tabular}
\end{table}

\begin{table}
\centering
\begin{footnotesize}
\caption[8]{\mbox{Initial parameters for benchmark point with $ {\boldsymbol{m_{16} = 25 \; {\rm TeV}}}$: }\\  (1/$\alpha_G, \, M_G, \, \epsilon_3$) = ($25.83,
\, 4.17 \times 10^{16}$ GeV, $\, -2.55$ \%), \makebox[1.8em]{ }\\
($\lambda, \, \lambda \epsilon, \, \sigma, \, \lambda \tilde \epsilon, \, \rho, \,
\lambda \epsilon', \, \lambda \epsilon \xi$) = ($ 0.61, \, 0.031, \, 1.17, \, 0.0049, \,
0.070, \, -0.0019, \, 0.0037  $),\\
($\Phi_\sigma, \, \Phi_{\tilde \epsilon}, \, \Phi_\rho, \,
\Phi_\xi$) =  ($0.513, \, 0.542, \, 3.969, \, 3.503$) rad,
\makebox[6.6em]{ }\\
($m_{16}, \, M_{1/2}, \, A_0, \, \mu(M_Z)$) = ($25000,\, 158, \,
-51365, \,
1348$) GeV,\\
($(m_{H_d}/m_{16})^2, \, (m_{H_u}/m_{16})^2, \, \tan\beta$) = ($1.86,  \, 1.61, \, 49.98$) \\
($M_{R_3}, \, M_{R_2}, \, M_{R_1}$) = ($3.2 \times 10^{13}$ GeV, $\,
 6.1 \times 10^{11} $ GeV, $\, 0.9 \times
10^{10} $ GeV) \label{t:fit25tev} 
}
\end{footnotesize}
\vspace{5mm}
\begin{tabular}{|l|l|l|l|l|}
\hline
Observable  &  Fit value  &  Exp. value  &  Pull & Sigma  \\
\hline
$M_Z$ &              91.1876         &  91.1876         &  0.0000          &  0.4559          \\
$M_W$ &              80.6192         &  80.3850         &  0.5824          &  0.4022          \\
$1/\alpha_{em}$ &    137.1624        &  137.0360        &  0.1844          &  0.6852          \\
$G_{\mu} \times 10^5$ & 1.1754          &  1.1664          &  0.7749          &  0.0117          \\
$\alpha_3$ &         0.1185          &  0.1184          &  0.0889          &  0.0009          \\
\hline
$M_t$ &              174.5241        &  173.5000        &  0.7735          &  1.3238          \\
$m_b(m_b)$ &         4.1789          &  4.1800          &  0.0307          &  0.0366          \\
$M_{\tau}$ &         1.7761          &  1.7768          &  0.0800          &  0.0089          \\
\hline
$m_c(m_c)$ &         1.2529          &  1.2750          &  0.8559          &  0.0258          \\
$m_s$ &              0.0963          &  0.0950          &  0.2652          &  0.0050          \\
$m_d/m_s$  &         0.0702          &  0.0526          &  3.1726          &  0.0055          \\
$1/Q^2$ &            0.0019          &  0.0019          &  0.3379          &  0.0001          \\
$M_{\mu}$ &          0.1057          &  0.1057          &  0.1533          &  0.0005          \\
$M_e \times 10^4$ &  5.1102          &  5.1100          &  0.0083          &  0.0255          \\
\hline
$|V_{us}|$ &         0.2244          &  0.2252          &  0.5407          &  0.0014          \\
$|V_{cb}|$ &         0.0411          &  0.0406          &  0.2090          &  0.0020          \\
$|V_{ub}| \times 10^3$ & 3.1806          &  3.7700          &  0.6933          &  0.8502          \\
$|V_{td}| \times 10^3$ & 8.9530          &  8.4000          &  0.9193          &  0.6015          \\
$|V_{ts}|$ &         0.0402          &  0.0429          &  1.0358          &  0.0026          \\
$\text{sin} 2\beta$ & 0.6318          &  0.6790          &  2.3268          &  0.0203          \\
\hline
$\epsilon_K$ &       0.0024          &  0.0022          &  0.8902          &  0.0002          \\
$\Delta M_{Bs}/\Delta M_{Bd}$ & 35.1576         &  35.0600         &  0.0139          &  7.0246          \\
$\Delta M_{Bd} \times 10^{13}$ & 4.1075          &  3.3370          &  1.1531          &  0.6682          \\
\hline
$m^2_{21} \times 10^5$ &   7.5325          &  7.5450          &  0.0229          &  0.5463          \\
$m^2_{31} \times 10^3$ &  2.4814          &  2.4800          &  0.0066          &  0.2104          \\
$\text{sin}^2 \theta_{12}$ & 0.2978          &  0.3050          &  0.2069          &  0.0350          \\
$\text{sin}^2 \theta_{23}$ & 0.5109          &  0.5050          &  0.0358          &  0.1650          \\
$\text{sin}^2 \theta_{13}$ & 0.0140          &  0.0230          &  1.2789          &  0.0070          \\
\hline
$M_h$ &              125.21          &  125.30          &  0.0293          &  3.0676          \\
\hline
$BR(B \rightarrow X_s \gamma) \times 10^4$ & 3.4074          &  3.4300          &  0.0138          &  1.6374          \\
$BR(B_s \rightarrow \mu^+ \mu^-) \times 10^9$ & 2.6112          &  3.2000          &  0.3610          &  1.6308          \\
$BR(B_d \rightarrow \mu^+ \mu^-) \times 10^{10}$ & 1.0779          &  8.1000          &  0.0000          &  5.2559          \\
$BR(B \rightarrow \tau \nu) \times 10^5$ & 6.4123          &  16.6000         &  1.1406          &  8.9320          \\
$BR(B \rightarrow K^*\mu^+ \mu^-)$(low) $\times 10^8$& 5.0511          &  19.7000         &  1.2202          &  12.0051         \\
$BR(B \rightarrow K^*\mu^+ \mu^-)$(high) $\times 10^8$ & 7.6223          &  12.0000         &  0.6010          &  7.2835          \\
$q_0^2(B \rightarrow K^* \mu^+ \mu^-)$ & 4.4839          &  4.9000          &  0.3198          &  1.3009          \\
\hline
\multicolumn{3}{|l}{Total $\chi^2$}  &  \textbf{27.9288}&  \\\hline
\end{tabular}
\end{table}

\begin{table}
\centering
\caption[8]{\mbox{Initial parameters for benchmark point with $ {\boldsymbol{m_{16} = 30 \; {\rm TeV}}}$: }\\  (1/$\alpha_G, \, M_G, \, \epsilon_3$) = ($25.86,
\, 4.36 \times 10^{16}$ GeV, $\, -2.81$ \%), \makebox[1.8em]{ }\\
($\lambda, \, \lambda \epsilon, \, \sigma, \, \lambda \tilde \epsilon, \, \rho, \,
\lambda \epsilon', \, \lambda \epsilon \xi$) = ($ 0.62, \, 0.031, \, 1.18, \, 0.0050, \,
0.069, \, -0.0020, \, 0.0037  $),\\
($\Phi_\sigma, \, \Phi_{\tilde \epsilon}, \, \Phi_\rho, \,
\Phi_\xi$) =  ($0.507, \, 0.534, \, 4.005, \, 3.514$) rad,
\makebox[6.6em]{ }\\
($m_{16}, \, M_{1/2}, \, A_0, \, \mu(M_Z)$) = ($30000,\, 161, \,
-61640, \,
1647$) GeV,\\
($(m_{H_d}/m_{16})^2, \, (m_{H_u}/m_{16})^2, \, \tan\beta$) = ($1.86,  \, 1.63, \, 50.15$) \\
($M_{R_3}, \, M_{R_2}, \, M_{R_1}$) = ($3.2 \times 10^{13}$ GeV, $\,
 6.4 \times 10^{11} $ GeV, $\, 0.9 \times
10^{10} $ GeV) 
} \label{t:fit30tev}
\vspace{5mm}
\begin{tabular}{|l|l|l|l|l|}
\hline
Observable  &  Fit value  &  Exp. value  &  Pull & Sigma  \\
\hline
$M_Z$ &              91.1876         &  91.1876         &  0.0000          &  0.4559          \\
$M_W$ &              80.6519         &  80.3850         &  0.6637          &  0.4022          \\
$1/\alpha_{em}$ &    137.0422        &  137.0360        &  0.0091          &  0.6852         \\
$G_{\mu} \times 10^5$ & 1.1785          &  1.1664          &  1.0410          &  0.0117        \\
$\alpha_3$ &         0.1187          &  0.1184          &  0.2850          &  0.0009          \\
\hline
$M_t$ &              175.0383        &  173.5000        &  1.1620          &  1.3238          \\
$m_b(m_b)$ &         4.1782          &  4.1800          &  0.0484          &  0.0366          \\
$M_{\tau}$ &         1.7764          &  1.7768          &  0.0419          &  0.0089          \\
\hline
$m_c(m_c)$ &         1.2504          &  1.2750          &  0.9540          &  0.0258          \\
$m_s$ &              0.0972          &  0.0950          &  0.4362          &  0.0050          \\
$m_d/m_s$  &         0.0712          &  0.0526          &  3.3536          &  0.0055          \\
$1/Q^2$ &            0.0019          &  0.0019          &  0.3029          &  0.0001          \\
$M_{\mu}$ &          0.1058          &  0.1057          &  0.1928          &  0.0005          \\
$M_e \times 10^4$ &  5.1097          &  5.1100          &  0.0125          &  0.0255          \\
\hline
$|V_{us}|$ &         0.2244          &  0.2252          &  0.5324          &  0.0014          \\
$|V_{cb}|$ &         0.0405          &  0.0406          &  0.0988          &  0.0020          \\
$|V_{ub}| \times 10^3$ & 3.1793          &  3.7700          &  0.6947          &  0.8502          \\
$|V_{td}| \times 10^3$ & 8.9071          &  8.4000          &  0.8431          &  0.6015          \\
$|V_{ts}|$ &         0.0396          &  0.0429          &  1.2682          &  0.0026          \\
$\text{sin} 2\beta$ & 0.6380          &  0.6790          &  2.0207          &  0.0203          \\
\hline
$\epsilon_K$ &       0.0022          &  0.0022          &  0.0877          &  0.0002          \\
$\Delta M_{Bs}/\Delta M_{Bd}$ & 34.0021         &  35.0600         &  0.1506          &  7.0246          \\
$\Delta M_{Bd} \times 10^{13}$ & 4.1018          &  3.3370          &  1.1445          &  0.6682          \\
\hline
$m^2_{21} \times 10^5$ &   7.5705          &  7.5450          &  0.0467          &  0.5463          \\
$m^2_{31} \times 10^3$ &  2.4783          &  2.4800          &  0.0081          &  0.2104          \\
$\text{sin}^2 \theta_{12}$ & 0.3057          &  0.3050          &  0.0213          &  0.0350          \\
$\text{sin}^2 \theta_{23}$ & 0.5036          &  0.5050          &  0.0087          &  0.1650          \\
$\text{sin}^2 \theta_{13}$ & 0.0130          &  0.0230          &  1.4280          &  0.0070          \\
\hline
$M_h$ &              125.88          &  125.30          &  0.1876          &  3.0676          \\
\hline
$BR(B \rightarrow X_s \gamma) \times 10^4$ & 3.3931          &  3.4300          &  0.0225          &  1.6374          \\
$BR(B_s \rightarrow \mu^+ \mu^-) \times 10^9$ & 2.6139          &  3.2000          &  0.3594          &  1.6308          \\
$BR(B_d \rightarrow \mu^+ \mu^-) \times 10^{10}$ & 1.0748          &  8.1000          &  0.0000          &  5.2559          \\
$BR(B \rightarrow \tau \nu) \times 10^5$ & 6.4081          &  16.6000         &  1.1411          &  8.9320          \\
$BR(B \rightarrow K^*\mu^+ \mu^-)$(low) $\times 10^8$& 4.9279          &  19.7000         &  1.2305          &  12.0051         \\
$BR(B \rightarrow K^*\mu^+ \mu^-)$(high) $\times 10^8$ & 7.4423          &  12.0000         &  0.6257          &  7.2835          \\
$q_0^2(B \rightarrow K^* \mu^+ \mu^-)$ & 4.4707          &  4.9000          &  0.3300          &  1.3009          \\
\hline
\multicolumn{3}{|l}{Total $\chi^2$}  &  \textbf{29.4783}&  \\
\hline\end{tabular}
\end{table}

\clearpage
\newpage

\bibliography{bibliography}

\bibliographystyle{utphys}

\end{document}